\documentclass[aip,graphicx,amsmath,amssymb,reprint]{revtex4-1}

\usepackage{graphicx}% Include figure files
\usepackage{dcolumn}% Align table columns on decimal point
\usepackage{bm}% bold math
\usepackage[outdir=./]{epstopdf}
\usepackage[utf8]{inputenc}
\usepackage[T1]{fontenc}
\usepackage{mathptmx}
\usepackage{etoolbox}

\usepackage{color}

\usepackage[caption=false]{subfig}

\usepackage[dvipsnames]{xcolor}
\usepackage{textcomp}

\draft % marks overfull lines with a black rule on the right

\makeatletter
\def\@email#1#2{%
 \endgroup
 \patchcmd{\titleblock@produce}
  {\frontmatter@RRAPformat}
  {\frontmatter@RRAPformat{\produce@RRAP{*#1\href{mailto:#2}{#2}}}\frontmatter@RRAPformat}
  {}{}
}%
\makeatother

\begin{document}

%\preprint{}

\title[Simulations of electrostatic waves in non-neutral plasmas]{Eulerian simulations of electrostatic waves in plasmas with a single sign of charge} %Title of paper


%\author{}
%\thanks{}
%\affiliation{}

\author{S. Cristofaro}
%\homepage{https://orcid.org/0000-0003-0437-2517}
\affiliation{Dipartimento di Fisica, Universit\`a della Calabria, I-87036 Rende (CS), Italy}

\author{O. Pezzi}
\email{oreste.pezzi@istp.cnr.it}
%\homepage{https://orcid.org/0000-0002-7638-1706}
\affiliation{Istituto per la Scienza e Tecnologia dei Plasmi (ISTP), Consiglio Nazionale delle Ricerche, Via Amendola 122/D, I-70126 Bari, Italy}

\author{T.M. O'Neil}
%\homepage{https://orcid.org/0000-0002-9904-1368}
\affiliation{Department of Physics, University of California at San Diego, La Jolla, California, 92093}

\author{P. Veltri}
%\homepage{https://orcid.org/0000-0002-7412-1660}
\affiliation{Dipartimento di Fisica, Universit\`a della Calabria, I-87036 Rende (CS), Italy}

\author{F. Valentini}
%\homepage{https://orcid.org/0000-0002-1296-1971}
\affiliation{Dipartimento di Fisica, Universit\`a della Calabria, I-87036 Rende (CS), Italy}

\date{\today}

\begin{abstract}
An Eulerian, numerical simulation is used to model the launching of plasma waves  in a non-neutral plasma that is confined in a Penning-Malmberg trap. The waves are launched by applying an oscillating potential to an electrically isolated sector at one end of the conducting cylinder that bounds the confinement region and are received by another electrically isolated sector at the other end of the cylinder. The launching of both Trivelpiece-Gould waves and electron acoustic waves is investigated. Adopting a stratagem, the simulation captures essential features of the finite length plasma, while retaining the numerical advantages of a simulation employing periodic spatial boundary conditions. As a benchmark test of the simulation, the results for launched Trivelpiece-Gould waves of small amplitude are successfully compared to a linearized analytic solution for these fluctuations.
\end{abstract}


\maketitle %\maketitle must follow title, authors, abstract and \pacs

%\section{}
%\label{}
%\subsection{}
%\subsubsection{}

% If in two-column mode, this environment will change to single-column format so that long equations can be displayed. 
%\begin{widetext}
%$$\mbox{put long equation here}$$
%\end{widetext}

%%%%%%%%%%%%%%%%%%%%%%%
\section{Introduction}\label{sect:intro}
%%%%%%%%%%%%%%%%%%%%%%%
Nowadays, the rapid increase in the capabilities of computational resources gives the possibility of running Eulerian simulations for the description of the kinetic dynamics of collisionless plasmas in physical conditions close to reality. Eulerian codes have then become in the last few years an indispensable tool to support the physical interpretation of the measurements both from space satellites and laboratory experiments. At variance with Lagrangian Particle In Cell (PIC) codes \citep{birdsall2004plasma}, in which the equations of motion of a large number of macro-particles are integrated in time under the effects of electric and magnetic fields, Eulerian codes \citep{mangeney2002numerical, valentini2005self,valentini2007hybrid,perrone2013vlasov} solve numerically the kinetic equation for the particle distribution function in phase space, self-consistently coupled to the Maxwell's equations for fields.

It is clear that the Eulerian algorithms are, in general, more computationally demanding than PIC algorithms in terms of both execution time and memory requirements for data allocation and storage. On the other hand, the main advantage of Eulerian schemes with respect to Lagrangian ones principally consists in the fact that the outcome of Eulerian computations does not suffer from the statistical noise intrinsic in Lagrangian schemes, which is mainly due to the limited number of macro-particles that can be loaded in a PIC simulation. Specifically, this noise can affect the phase-space features as well as the higher-order moments of the particle distribution function \citep{camporeale2011dissipation,pezzi2017colliding,bacchini2022kinetic}.

In recent years, the number of numerical analyses based on the Eulerian integration of the kinetic equations has impressively increased, especially in the study of the kinetic dynamics of the turbulent solar wind \citep{valentini2016differential,cerri2017kinetic}. Moreover, huge efforts have been also devoted to modeling weakly-collisional plasmas as the interplanetary medium (e.g., Refs. \citep{pezzi2015collisional, pezzi2019protonproton} and references therein), where even a low collisionality could play a relevant role in the processes of turbulent energy dissipation and plasma heating. The inclusion of collisional effects might be even more significant for laboratory plasmas\citep{affolter2016first,anderegg2017measurements}.

\begin{figure}
\includegraphics[width=\columnwidth]{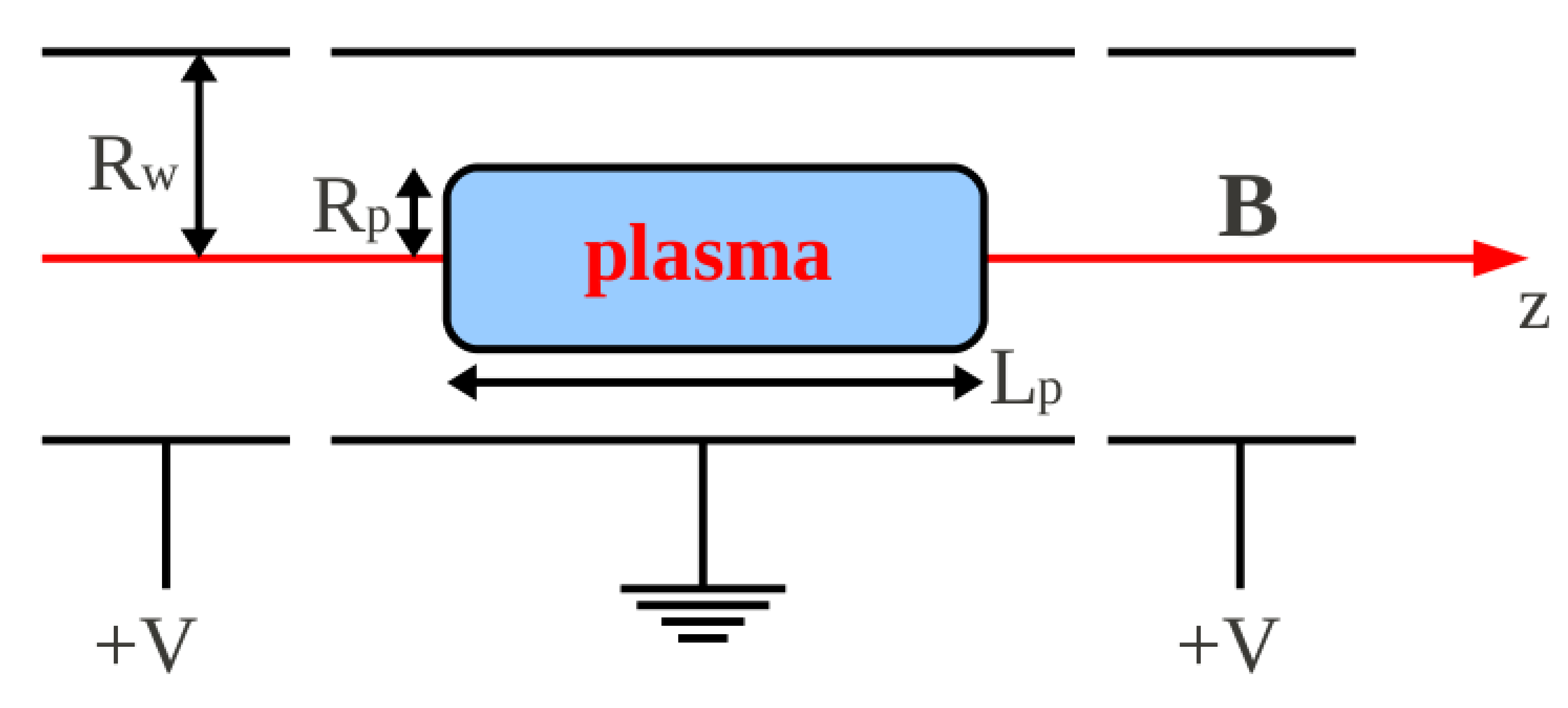}
\caption{A schematic representation of a Penning-Malmberg apparatus.} \label{fig:trap}                             % FIGURE 1
\end{figure}

Here, we use an Eulerian simulation to model the launching of plasma waves in a nonneutral plasma confined in a Penning-Malmberg trap, whose schematic representation is given in Fig. \ref{fig:trap}. Nonneutral plasmas confined in these devices show a rich collective behavior akin to quasi-neutral plasmas. Their collisionality is often quite small, thus allowing kinetic effects to play a significant role in the plasma dynamics. Several fluctuations and instabilities are supported by these plasmas. Trivelpiece-Gould waves (TGWs)\citep{trivelpiece1959space}—the counterpart of Langmuir waves in nonneutral plasmas, Electron Acoustic waves (EAWs) \citep{anderegg2009waveparticle,anderegg2009electron,valentini2012undamped} are typical longitudinal fluctuations occurring in Penning-Malmberg devices, while diocotron\citep{kabantsev2014diocotron,chim2016diocotron,chim2021diocotron}, cyclotron\citep{dubin2013cyclotron,affolter2015cyclotron} and Bernstein\citep{walsh2018bernstein} waves oscillate in the transverse directions. The perpendicular dynamics shows also significant analogies with a two-dimensional inviscid fluid\citep{driscoll1990experiments}, thus motivating several experimental and numerical efforts to study, for example, the vortex dynamics subject to an external strain \citep{hurst2016evolution,hurst2021adiabatic}. More recently Penning-Malmberg devices have been exploited in experiments aimed at determining the gravitational properties of antimatter\citep{perez2015gbar}. In our numerical model, waves are excited by applying an oscillating potential to an electrically isolated section of the conducting cylindrical wall that bounds the confinement region.  This launching electrode is near one end of the cylinder, and a receiving electrode is located near the other end. As will be described, an artifice is used to capture essential aspects of launching in a finite length plasma, while retaining the numerical advantages of periodic boundary conditions on an infinitely long plasma column.   

The launching of both Trivelpiece-Gould Waves (TGWs) \citep{trivelpiece1959space} and Electron Acoustic waves (EAWs) is investigated. TGWs are simply Langmuir waves on a long magnetized plasma column, while EAWs are intrinsically nonlinear waves with phase velocity near the thermal velocity. These waves were predicted in 1991 by \citet{holloway1991undamped} for the case of an infinite homogeneous neutral plasma. By solving for the roots of the electrostatic dielectric function in the presence of an infinitesimal plateau at $v_\phi=\omega/k$, it has been shown that undamped plasma oscillations can be obtained with wavenumber and real part of the wave frequency lying on the so-called thumb curve\citep{holloway1991undamped}. In the original theory, the ions do not participate in the wave motion but simply provide a uniform neutralizing background charge, so a close analogue of such waves exist in nonneutral plasmas. Indeed, for EAWs, Landau damping \citep{landau1946vibration} is inhibited by the presence of a flat region (plateau) in the particle velocity distribution at the wave phase speed, produced by the trapping of particles \citep{oneil1965collisionless} in the wave potential well.

Previous PIC \citep{valentini2006excitation} and Eulerian simulations \citep{afeyan2004kinetic,johnston2009persistent,valentini2012undamped} have successfully excited EAWs. Moreover, experimental results\citep{anderegg2009waveparticle,anderegg2009electron} confirmed the existence of the EAW branch and also suggested that the excitation of undamped electrostatic waves can be obtained even off of the usual thumb curve\citep{holloway1991undamped}. This latter evidence has been further confirmed by the theoretical and numerical results\citep{valentini2012undamped}. In both numerical simulations and laboratory experiments, the plateau needed to turn off Landau damping for the excitation of the EAWs is generated through an external driving electric field applied on the plasma initially at equilibrium. 

However, in the previous simulations of EAWs and TGWs, the launched wave was selected by wave number matching. The driving electric field was sinusoidal with an axial wave number chosen to match that of the wave to be launched \citep{valentini2012undamped}. In some simulations~\citep{valentini2011new,valentini2011excitation}, a radial wave number, again chosen to match that of the launched wave, was included in an ad hoc manner in the mode equation. The launching was most effective when the frequency of the driver was chosen to be near that of the wave to be launched, but the particular triggered wave was selected by wave-number matching. Because of the sinusoidal profile of the driving electric field, the driver was spatially orthogonal to all of the modes except the launched one. In contrast, for the simulations discussed here and for the experiments, the excited fluctuations are selected by frequency matching. The driver field is localized in space and is not characterized by any axial or radial mode numbers of a wave to be launched. The spatial structure of the launched waves is determined by the plasma itself, rather than by the spatial structure of the launching field. 

\begin{figure}
\includegraphics[width=\columnwidth]{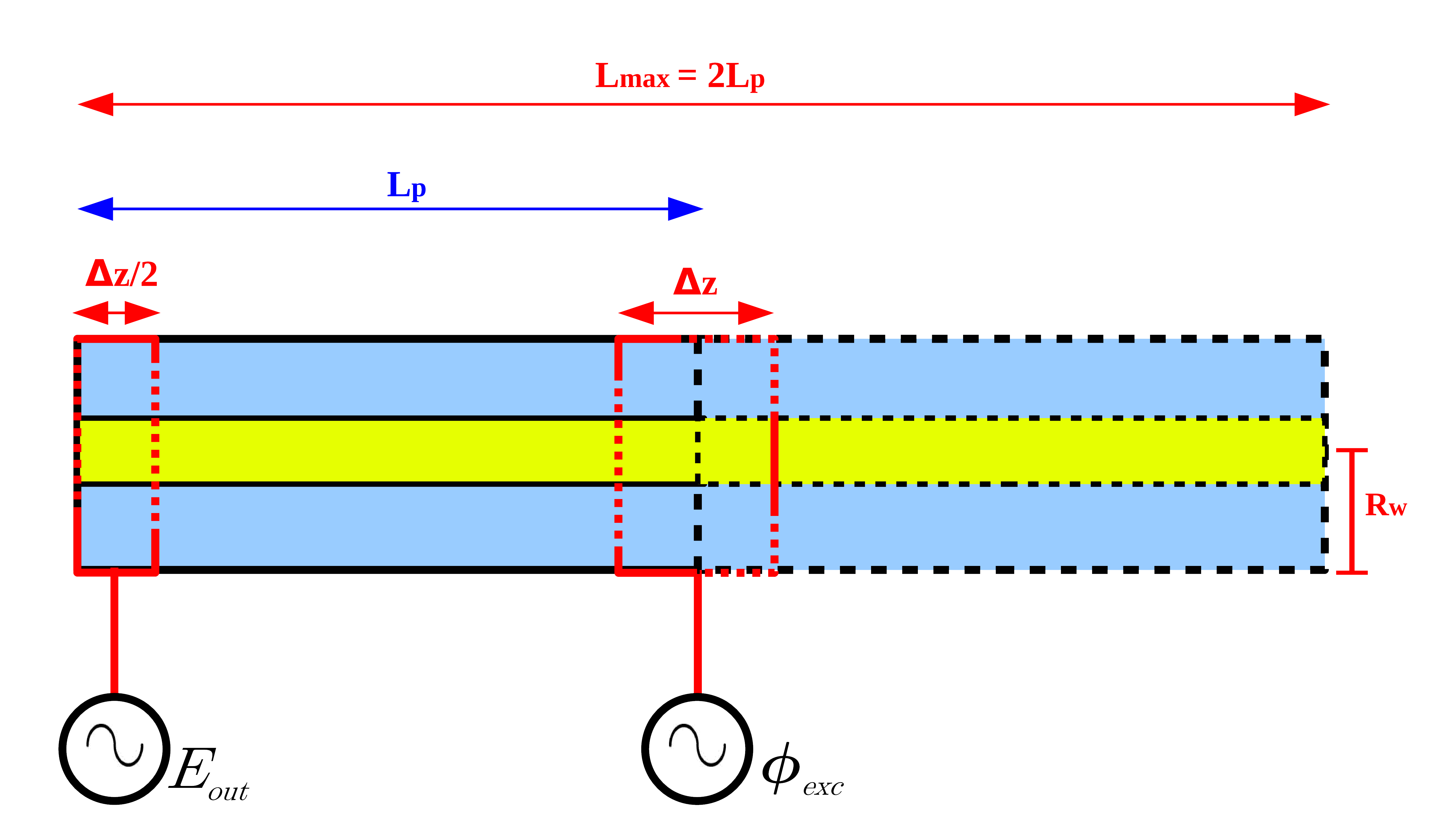}
\caption{Representation of the $(r,z)$ numerical domain for the Eulerian simulations.} \label{fig:picnum}          % FIGURE 2
\end{figure}

We refer to the frequency selection process as adiabatic frequency selection. The oscillating driving potential applied to the wall electrode is typically of the form $V(t)=V_D h(t)\sin(\omega_{D}t)$, where $\omega_{D}$ is the driver frequency, $V_D$ is the driver amplitude, and $h(t)$ is a factor that smoothly turns on and off the oscillating drive potential. Typically, $h(t)$ rises smoothly from zero to the unit value on a timescale $\Delta t$, and then, after some time, smoothly returns to zero, again on the timescale $\Delta t$. For a mode characterized by frequency $\omega$, being $\left| \omega-\omega_{D}\right|\Delta t > 1$, we will see that the mode amplitude adiabatically follows $h(t)$ as it increases, but also adiabatically follows $h(t)$ back down as it returns to zero, that is, the wave does not launch and ring after the drive potential is turned off. In contrast, when  $\left| \omega-\omega_{D}\right|\Delta t \lesssim 1$, $h(t)$ is not slowly varying compared to the frequency difference, and the mode does launch, ringing after the drive potential is turned off. This type of drive potential selects modes that differ in frequency from the drive frequency by order $1 /\Delta t$.

For the case of a linear TGW, the launching process can be described analytically, and demonstrate the adiabatic frequency selection process. The analytic results are used to benchmark the numerical results for launching of a small-amplitude TGW. The same cannot be done for EAWs, since the launching process for these waves involves nonlinear processes.

The Eulerian simulation here presented solves the coupled drift kinetic equation and Poisson’s equation for the simple case where the plasma remains azimuthally symmetric. In our model, we assume a top-hat profile in the radial direction for the equilibrium plasma density. This approximation neglects the sheath at the plasma radius, where the density decreases exponentially to zero. However, our model is able to capture essential features of a finite length bounded plasma, while maintaining the numerical advantages of periodic boundary conditions on a long column. In particular, the plasma column is assumed to have flat ends where the particles undergo specular reflection. On the other hand, in real experiments, plasma ends are rounded and the reflection occurs on the scale length of a few Debye lengths. However, our assumption of specular reflection from flat ends is reasonable when the axial wavelength is long compared to the scale length of the rounding and of the Debye length.

When a mode is excited in a real plasma, the mode potential extends somewhat beyond the end into the vacuum, and both the value of the potential and the value of the normal component of the mode displacement current must be continuous across the plasma end. The displacement current in the plasma is the product of the plasma dielectric times the normal component of the mode electric field, but in the vacuum beyond the end, the displacement current is simply the normal component of the mode electric field. Thus, when the mode frequency is small compared to the plasma frequency and, hence, the magnitude of the plasma dielectric is much larger than unity $\left|1-\omega_{p}^{2}/\omega^{2}\right|\gg 1$, the matching across the plasma ends requires the normal component of the mode electric field in the plasma to be near zero.  If one end of the flat-ended model plasma column is at $z=0$ and the other end is at $z=L_{p}$, where $L_{p}$ is the plasma length, we require that the axial component of the mode electric field vanish at both $z=0$ and at $z=L_{p}$.

To conveniently model specular reflection of particles at the plasma ends, we create a mirror plasma and potential in the domain $z=L_{p}$ to $z=2L_{p}$ by reflection of the real plasma and potential about the end $z=L_{p}$. This is shown in Fig. \ref{fig:picnum}. The domain of the model plasma then extends from $z=0$ to $z=2L_{p}$, with periodic boundary conditions imposed on the potential at $z=0$ and $z=2L_{p}$. The plasma potential is even under axial reflection about both $z=0$ and $z=L_{p}$. In this way, the simulation models specular reflection at the flat ends of the plasma.  

Nevertheless, this model does not capture all aspects of mode reflection in a real plasma. We often think of a standing wave on a string as being composed of two traveling waves propagating in opposite directions and reflecting into one another at the ends of the string. This picture also describes standing modes on a finite length plasma column, providing that the plasma density extends uniformly out to the wall~\citep{prasad1983waves,anderson2011degenerate}. In this case, the radial dependence of the mode potential in the plasma and in the vacuum beyond the plasma end are the same, and a given traveling wave reflects only into itself. More generally, when the plasma density does not extend uniformly out to the wall, the radial dependence of the wave potential in the plasma and in the vacuum are not the same, and, as a result, a given traveling wave reflects into many different waves~ \citep{prasad1983waves,anderson2011degenerate}. The mode vacuum potential drives the other modes at the frequency of the incident mode, so the reflection-produced mode mixing is important only when there are nearly degenerate low-order modes in the frequency range of the launcher, and our simple model neglects the reflection mixing.   

Another way to understand this approximation is to first observe that there is no reflection mixing for true eigenmodes of the finite length plasma; that is what is meant by an eigenmode. The trouble is that a standing wave on a long column characterized by single axial and radial wave numbers, is not a true eigenmode of the finite length plasma. Reflection mixing couples in small admixtures of other modes and produces a small correction to the frequency. Our simple model neglects these small corrections. In defense of the simple model, we note that it has done a good job in predicting the frequencies and damping rates of plasma modes in many experiments with finite length nonneutral plasmas in Penning-Malmberg traps~\citep{afeyan2004kinetic,valentini2006excitation,valentini2011new,valentini2011excitation,valentini2012undamped,affolter2018trapped,affolter2019fluid}.

The paper is organized as follows. Sections \ref{sect:wavedisp}- \ref{sect:Launching_theory} provide analytic backup for the simulations in Sec. \ref{sect:numsim}. Section \ref{sect:wavedisp} discusses linear dispersion theory for TGWs in our model plasma, finding a simple translation between the TGW dispersion for say the axial and radial mode $(n,m)$ and dispersion theory for plasma modes an infinite homogeneous plasma. Section \ref{sect:launching} discusses the usual method of launching waves in simulations, that is, wave evolution from an initial perturbation in the plasma. The initial density perturbation is projected onto the set of eigenmodes for the TGWs, and the orthogonality of the eigenmodes insures that a given mode evolves only from its own projection of the initial density perturbation. The mode selection is through the spatial eigenmode structure. Section \ref{sect:Launching_theory} discusses the launching of TGWs from an oscillating potential applied to a wall electrodes, and demonstrates the adiabatic frequency selection of modes. In Section \ref{sect:nummodel}, we describe the numerical method employed for launching waves in an azimuthally symmetric non-neutral plasma column. Then, in Section \ref{sect:numsim} we present the results of Eulerian simulations for the excitation of TG and EA fluctuations. Finally, a summary and the conclusions are given in Section \ref{sect:concl}.

%%%%%%%%%%%%%%%%%%%%%%%%%%%%%%%%%%%%%
\section{Linear dispersion theory for TGWs}\label{sect:wavedisp}
%%%%%%%%%%%%%%%%%%%%%%%%%%%%%%%%%%%%%
We consider a cylindrical pure ion plasma column of uniform density $n_0$ as shown in Fig. \ref{fig:trap}. Let $(r,\theta, z)$ be a cylindrical coordinate system with the $z-$axis coincident with the axis of the plasma column and the plane $z=0$ coincident with one end of the plasma. The plasma is assumed to have flat ends and to be of length $L_{p}$ and radius $R_{p}$. The plasma resides in a conducting cylinder of radius $R_{w}$, where $R_{p} < R_{w} \ll L_{p}$. We consider a plasma consisting of singly-ionized positive ions \citep{anderegg1997new,affolter2019fluid} of charge $e$ and mass $m$. The radial confinement is provided by a uniform axial magnetic field $B_0$ and the axial confinement by positive end potentials. The magnetic field is sufficiently strong that the cyclotron frequency $\Omega_{c}=eB_0/mc$ is large compared to the plasma frequency $\omega_{p}=\sqrt{4\pi n_0 e^2/m}$, which also implies that the cyclotron radius $r_{c}=v_{th}/\Omega_c$ is small compared to the Debye length $\lambda_{D}=\sqrt{k_B T/4\pi n_0 e^2}$, where $v_{th}=\sqrt{k_B T/m}$ is the thermal speed and $T$ the plasma temperature. For the collection of charges to qualify as a plasma, the Debye length must be small compared to the plasma dimensions, so the overall spatial ordering is given by the inequalities $r_{c} \ll \lambda_{D} \ll R_{p} < R_{w} \ll L_{p}$. We will consider modes with frequencies that are small compared to the plasma frequency, but large compared to the collision frequency $\nu$, so the overall frequency ordering is $\nu \ll \omega \ll \omega_{p} \ll \Omega_{c}$. 

This ordering allows a simplified description of the particle dynamics. At the end of the plasma, the confinement potential rises sharply on the scale of the Debye length, so the ordering $\lambda_{D} \ll L_{p}$ together with the limitation of mode wave lengths to be of order $L_{p}$ allows particle reflection at the column ends to be approximately specular. Also, the ion dynamics is well approximated by the drift kinetic equation, which for the case where the mode has no azimuthal dependence reduces to the simple form
\begin{equation}\label{eq:vla}
\frac{\partial F}{\partial t}+v\frac{\partial F}{\partial z}-\frac{e}{m}\frac{\partial\phi}{\partial z}\frac{\partial 
F}{\partial v}=0,
\end{equation}
where $F=F(r,z,v,t)$ is the particle distribution function and $\phi=\phi(r,z,t)$ the electric potential. We will find it convenient to write $F$ in the form $F=n_{0}(r)f(r,v,z,t)$, where the unperturbed density, $n_{0}(r)$, has the constant value $n_{0}$ for $r<R_{p}$ and the value zero for $R_{p}<r<R_{w}$, that is, the density has a top-hat profile. Such a choice does not model the sheath at $R_p$ \citep{prasad1979finite}, which is the region where the density drops exponentially to zero and whose dimension is only a few Debye lengths. Since the radial variable $r$ enters Eq. (\ref{eq:vla}) only parametrically, the factor $n_{0}(r)$ can be canceled out of all terms, and $F$ replaced by $f$. Equation (\ref{eq:vla}) must be solved in parallel with Poisson’s equation
\begin{equation}\label{eq:poi}
\frac 1 r \frac{\partial}{\partial r}r\frac{\partial\phi}{\partial r}+\frac{\partial^2\phi}{\partial z^2}=-4\pi 
en_{0}(r)\int dv \, f. 
\end{equation}
Within linear theory, the particle distribution function and potential are written in the forms $f=f_{0}(v)+\delta f(r,v,z,t)$ and $\phi=\phi_{0}(r)+\delta \phi(r,z,t)$, where $f_{0}(v)$ is such that $\int dv f_0(v)=1$, while $\delta f(r,v,z,t)$ and $\delta \phi(r,z,t)$ are first-order perturbations satisfying
\begin{equation}\label{eq:vla_linearized}
    \left(\dfrac{\partial}{\partial t} + v\dfrac{\partial}{\partial z}\right)\delta f = \dfrac{e}{m}\dfrac{\partial \delta \phi}{\partial z}\dfrac{\partial f_{0}}{\partial v},
\end{equation}
\begin{equation}\label{eq:poi_linearized}
   \left(\frac 1 r \frac{\partial}{\partial r}r\frac{\partial}{\partial r}+\frac{\partial^2}{\partial z^2}\right) \delta \phi = -4\pi e n_{0}(r)\int dv \, \delta f .
\end{equation}
Since particles bounce back and forth between the ends of the plasma column, it is necessary that $f_{0}(-v)=f_{0}(v)$. We will take $f_{0}(v)$ to have the form
\begin{equation}
    f_{0}(v) =  \dfrac{1}{\sqrt{2 \pi v_{th}^{2}}}\exp\left[\dfrac{-v^{2}}{2v_{th}^{2}}\right],
\end{equation}
which clearly satisfies this property.

Following the simple theory model described in Sec. \ref{sect:intro}, we extend the plasma to a length $2L_{p}$, as shown in Fig. \ref{fig:picnum}, impose periodic boundary conditions with periodicity $2L_{p}$, and require that $\partial \delta \phi / \partial z$ be zero at $z=0$ and $z=L_{p}$. This simple model mimics the specular reflection of particles at the ends of the real plasma. The perturbations can be written in the separable form
\begin{equation}\label{eq:f_pert}
    \delta f(r,z,v,t)\simeq\delta \hat{f}_k(r,v,p)\cos(kz)e^{pt},
\end{equation}
\begin{equation}\label{eq:phi_pert}
    \delta \phi(r,z,t)\simeq\delta \hat{\phi}_k(r,p)\cos(kz)e^{pt}.
\end{equation}
where both the periodicity and the requirement that $\partial \delta \phi/\partial z$ vanish at the plasma ends are satisfied when $k=k_{n}\equiv 2\pi n/{2L_{p}}=\pi n/{L_{p}}$, and $n$ is an integer. Note that in Eqs. (\ref{eq:f_pert}) and (\ref{eq:phi_pert}), by writing the time dependence as $e^{pt}$, we have anticipated the use of the Laplace transform in the discussion of wave launching. We will find solutions $p=p_{n,m}=i\omega_{n,m}+\gamma_{n,m}$ for modes  labeled by the axial and radial indices $(n,m)$, where $\omega_{n,m}$ is the mode frequency and $\gamma_{n,m}$ the damping rate. Substituting Eqs. (\ref{eq:f_pert}) and (\ref{eq:phi_pert}) into Eq. (\ref{eq:vla_linearized}) yields the solution
\begin{small}
 \begin{equation}
    \delta \hat{f}_{k_n}(r,v,p)=\dfrac{\dfrac{e}{m}ik_{n}\delta \hat{\phi}_{k_n}(r,p)\dfrac{\partial f_{0}(v)}{\partial v}}{2(ik_{n}v+p)} +\dfrac{-\dfrac{e}{m}ik_{n}\delta \hat{\phi}_{k_n}(r,p)\dfrac{\partial f_{0}(v)}{\partial v}}{2(-ik_{n}v+p)},
 \end{equation}
\end{small}
where the function $\cos(k_{n}z)$ has been written out in terms of complex exponentials to facilitate the solution. Substituting this expression into the right-hand side of Eq. (\ref{eq:poi_linearized}) and using the relation $f_{0}(-v)=f_{0}(v)$ to rewrite the second velocity integral yields the equation for the mode potential $\delta \hat{\phi}_{k_n}(r,p)$,
\begin{equation}
    \left[\dfrac{1}{r}\dfrac{\partial}{\partial r}r\dfrac{\partial}{\partial r} - k_{n}^{2}+\dfrac{n_{0}(r)}{n_{0}}K^{2}(k_{n},p)\right]\delta \hat{\phi}_{k_n}(r,p)=0,
\end{equation}
where the quantity $K^{2}$ is given by the expression
\begin{equation}\label{eq:Kappa2}
    K^{2}(k_{n},p) = \omega_{p}^{2}\int_{L} dv \dfrac{\dfrac{\partial f_0(v)}{\partial v}}{p/ik_{n}+v},
\end{equation}
and the $L$ on the integral sign indicates that the Landau contour is to be followed \citep{landau1946vibration}. In the absence of wave launching electrodes, the mode potential $\delta \hat{\phi}_{k_n}(r,p)$ must vanish on the conducting cylindrical wall at $r=R_{w}$. Ultimately, we will consider a case where there is an electrically isolated launching electrode, that is, a section of the wall where the oscillating potential is nonzero. However, as discussed below, that will be handled by a separate, launching potential. Here, we specify that the mode potential vanish everywhere on the conducting wall. Also, its radial derivative must vanish at $r=0$, while both the mode potential and its radial derivative must be continuous across the density discontinuity at $r=R_{p}$. 

Defining a transverse wave number through the relation where $k_{\perp}^{2}=K^{2}-k_{n}^{2}$, the expressions
\begin{widetext}
\begin{equation}\label{eq:psi}
  %\delta\phi(r) = 
  \delta \hat{\phi}_{k_n}(r,p) \propto \psi_{n,m}(r) = 
\begin{cases}
J_{0}(k_{\perp}r)\;\;\;\;\;\;\;\;\;\;\;\; &0 <r\leq R_p \\
J_{0}(k_{\perp}R_{p})\dfrac{I_{0}(k_{n}r)K_{0}(k_{n} R_{w})-I_{0}(k_{n}R_{w})K_{0}(k_{n}r)}{I_{0}(k_{n} R_{p})K_{0}(k_{n} R_{w})-I_{0}(k_{n} R_{w})K_{0}(k_{n}R_{p})}\;\; &R_p<r<R_w
\end{cases}
\end{equation}
\end{widetext}
satisfy the equations and all of the boundary conditions except continuity of $\partial \delta\phi/\partial r$ at $r=R_{p}$. Implementing the last boundary condition yields the requirement
\begin{small}
\begin{equation}\label{eq:phi_continuity}
    \dfrac{k_{\perp}J_{1}(k_{\perp}R_{p})}{J_{0}(k_{\perp}R_{p})} + \dfrac{k_{n}[I_{1}(k_{n}R_{p})K_{0}(k_{n}R_{w})+I_{0}(k_{n}R_{w})K_{1}(k_{n}R_{p})]}{I_{0}(k_{n}R_{p})K_{0}(k_{n}R_{w})-I_{0}(k_{n}R_{w})K_{0}(k_{n}R_{p})} = 0.
\end{equation}
\end{small}
For a given value of $k=k_n=\pi n/L_{p}$, Eq. (\ref{eq:phi_continuity}) determines a sequence of values $k_{\perp}=k_{\perp n,m}$, where $n$ and $m$ are respectively the longitudinal and radial quantum numbers for the eigenmode $\psi_{n,m}(r)$. Writing the mode equation in the form
\begin{equation}\label{eq:mode_equation_kn+knm}
    \left[\dfrac{1}{r}\dfrac{\partial}{\partial r}r\dfrac{\partial}{\partial r} -k_{n}^{2}+\dfrac{n_{0}(r)}{n_{0}}(k_{\perp n,m}^{2}+k_{n}^{2})\right]\psi_{n,m}(r)=0 \, ,
\end{equation}
multiplying by $\psi_{n,m'}(r)$, subtracting the resulting equation with $m$ and $m'$ interchanged, and integrating over $r dr$ shows that the radial eigenmodes $\psi_{n,m}(r)$ and $\psi_{n,m'}(r)$ satisfy the orthogonality property:
\begin{equation}
    (k_{\perp n,m}^{2}-k_{\perp n,m'}^{2})\int_{0}^{R_{w}}dr\,r \dfrac{n_{0}(r)}{n_{0}}\psi_{n,m}(r)\psi_{n,m'}(r)=0 \, .
\end{equation}
Thus, we may write the relation
\begin{eqnarray}
    \int_{0}^{R_{w}}dr\,r &&\dfrac{n_{0}(r)}{n_{0}}\psi_{n,m}(r)\psi_{n,m'}(r)= \nonumber \\ 
    &&\delta_{m,m'}\int_{0}^{R_{w}}dr\,r \dfrac{n_{0}(r)}{n_{0}}\psi_{n,m}(r)\psi_{n,m'}(r)
\end{eqnarray}

where $\delta_{m,m'}$ is a Kronecker delta.  Finally, we note that the full spatial eigenmodes
\begin{equation}
    \Psi_{n,m}(r,z)=\cos(k_{n}z)\psi_{n,m}(r)
\end{equation}
satisfy the orthogonality condition
\begin{eqnarray}
    \int_{0}^{2L_{p}} dz &&\int_{0}^{R_{w}} dr\,r \dfrac{n_{0}(r)}{n_{0}} \Psi_{n,m}(r,z)\Psi_{n',m'}(r,z) = \nonumber \\ &&\delta_{n,n'}\delta_{m,m'}L_{p}\int_{0}^{R_{w}}dr\,r \dfrac{n_{0}(r)}{n_{0}}\psi_{n,m}(r)\psi_{n,m'}(r).
\end{eqnarray}
The eigenfunctions $\cos(k_{n}z)$ are complete on the interval $(0,2L_{p})$ for the set of functions with zero derivative at $z=0, L_{p}$, and $z=2L_{p}$ and the set of functions $\psi_{n,m}(r)$ are complete on the radial domain of the plasma, $r=0$ to $r=R_{p}$, since the set is eigenfunction of a Sturm-Liouville equation~\citep{prasad1983waves,prasad1984vlasov}. The frequency corresponding to the $(n, m)$ eigenmode is determined by the dispersion relation
\begin{equation}
\label{eq:dispersion relation}
    0=k_{\perp n,m}^{2}+k_{n}^{2}-K^{2}(k_{n},p)=k_{n}^{2}D_{n,m}(k_{n},p),
\end{equation}
where 
\begin{equation}\label{eq:dielectric function}
    D_{n,m}(k_{n},p)=1+\dfrac{k_{\perp n,m}^{2}}{k_{n}^{2}}-\dfrac{\omega_{p}^{2}}{k_{n}^{2}}\int_{L}dv\dfrac{\dfrac{\partial f_0(v)}{\partial v}}{v+p/ik_{n}}
\end{equation}
is the effective plasma dielectric for the  eigenmode. The dispersion relation in Eq. (\ref{eq:dielectric function}) can be rewritten as
\begin{equation}\label{eq:TG_dispersion_relation_rewritten}
    1-\dfrac{\omega_{p}^{2}(n,m)}{k_{n}^{2}} \int_{L} dv \dfrac{\dfrac{\partial f_0(v)}{\partial v}}{v+p/ik_{n}} = 0,
\end{equation}
where $\omega_{p}^{2}(n,m)=(k_{n}^{2}\omega_{p}^{2})/(k_{n}^{2}+k_{\perp n,m}^{2})$ can be thought of as the effective plasma frequency for the mode $(n, m)$. This form of the dispersion relation is just the same as that for a Langmuir wave in an infinite homogeneous plasma, except that $\omega_{p}^{2}(n,m)$ has replaced $\omega_{p}^{2}$. For example, for the case of a cold plasma (i.e., $v_{th}^{2}\equiv k_B T/m = 0$) where the velocity integral reduces to the ratio $(-k_{n}^{2}/p^{2})$ and the plasma dielectric to $[1+\omega_{p}^{2}(n,m)/p^{2}]$ , the dielectric vanishes for $p_{n,m}=i\omega_{p}(n,m)$, that corresponds to the mode frequency $\omega_{n,m}\equiv \omega_{p}(n,m)$ and damping rate $\gamma_{n,m}=0$.

For a warm plasma with a Maxwellian equilibrium distribution function the thermal velocity $v_{th}=\sqrt{k_B T/m}$ enters the velocity integral, but is often replaced in formulas by the Debye length $\lambda_{D}=v_{th}/\omega_{p}$. Since the translation pursued here requires all plasma frequencies to be replaced by the effective plasma frequency $\omega_{p}(n,m)$, the Debye length must be replaced by the effective Debye length $\lambda_{D}(n,m)\equiv v_{th}/\omega_{p}(n,m)=\lambda_{D}\omega_{p}/\omega_{p}(n,m)$  for the mode $(n,m)$.

For the case where the wave phase speed is large compared to the thermal velocity and the Landau damping is weak, the translated Landau results \citep{landau1946vibration} are the following
\begin{eqnarray}
 \omega_{n,m} &&\simeq \omega_{p}(n,m)\left[1+\dfrac{3}{2}k_{n}^{2}\lambda_{D}^{2}(n,m)\right], \label{eq:real_frequency} \\
\gamma_{n,m} &&\simeq \dfrac{\pi}{2}\dfrac{\omega_{n,m}^{3}}{k_{n}\lvert k_{n}\rvert}\dfrac{\partial f_{0}}{\partial v} \Big|_{\frac{\omega_{n,m}}{k_{n}}}  \nonumber \\
&&\simeq-\sqrt{\dfrac{\pi}{8}} \dfrac{\omega_{p}(n,m)}{\lvert k_{n}^{3}\lambda_{D}^{3}(n,m)\rvert}exp\left(\dfrac{-1}{2k_{n}^{2}\lambda_{D}^{2}(n,m)}-\dfrac{3}{2}\right). \label{eq:imaginary_frequency}
\end{eqnarray}
We remark that, although this last form for the damping rate in Eq. \ref{eq:imaginary_frequency} is often quoted, it is less accurate for moderate damping than the previous form \citep{malmberg1964collisionless}, since it uses the Bohm-Gross approximation to evaluate the wave phase velocity in the exponential, and this approximation is not accurate enough when the phase velocity is large enough to handle moderate damping. For more accurate results, the plasma dispersion relation
\begin{equation}
  1+\dfrac{1}{k_{n}^{2}\lambda_{D}^{2}(n,m)}+\dfrac{\zeta}{k_{n}^{2}\lambda_{D}^{2}(n,m)}Z(\zeta) = 0
\end{equation}
should be solved numerically. Here, $\zeta \equiv i p/{(\sqrt{2}\omega_{p}(n,m)k_{n}\lambda_{D}(n,m))}$ is the scaled complex frequency, and $Z(\zeta)=\int_{L}dx e^{-x^{2}}/(x-\zeta)$ is the plasma dispersion function \citep{huba1998nrl}.

The ratio $\omega_{p}(n,m)/\omega_{p}$ determines the transformation between the solutions for an infinite homogeneous plasma and the $(n, m)$ mode in the bounded plasma, and  the allowed values for this ratio depend only on the dimensions $L_{p},R_{p}$ and $R_{w}$, and not on the equilibrium velocity distribution. The frequency  $\omega_{n,m}$ is an increasing function of $n$ and a decreasing function of $m$. For a warm plasma, the ratio of the phase velocity to the thermal velocity decreases as $m$ increases, so the Landau damping increases.

In Sec. \ref{sect:intro}, we described a caveat to our simple model concerning the reflection mixing of nearly degenerate modes in a real plasma. In determining the importance of this effect, one must first remember that the real plasma is invariant under axial reflection about the plasma mid-plane $z=L_{p}/2$, so axially the real modes have either odd parity or even parity. Therefore, mixing can occur only between modes of even $n$, or separately between modes of odd $n$. As shown in Appendix, an estimate of reflection mixing based on \citet{anderson2011degenerate} indicates that the $(n,m)$ has a small admixture of a nearly degenerate mode $(n^\prime,m^\prime)$ given approximately by the amplitude ratio
\begin{equation}\label{eq:amplitude_ratio}
    \dfrac{A_{n',m'}}{A_{n,m}} \sim  \dfrac{R_{w}R_{p}}{L_{p}^{2}\,n'\pi(1+3m')}\dfrac{\omega_{p}}{\lvert\omega_{n,m}-\omega_{n',m'}\rvert} .
\end{equation}
This general expression provides the admixture of the $(n', m')$ mode with the main $(n, m)$ mode. For a sufficiently cold plasma, there always are higher order $m'$ modes with close enough degeneracy to a low-order mode that there is substantial mixing. However, for a warm plasma, Landau damping of the higher $m'$ modes spoils the degeneracy. Note also that, for a warm plasma, the frequency difference in the denominator of Eq. (\ref{eq:amplitude_ratio}) includes both real and imaginary parts.

%%%%%%%%%%%%%%%%%%%%%%%%%%%%%%%%%%%%%%
\section{Initial value problem for Trivelpiece-Gould waves}\label{sect:launching}
%%%%%%%%%%%%%%%%%%%%%%%%%%%%%%%%%%%%%%

In most simulations, waves evolve from an initial perturbation, and the particular wave or waves excited are specified by the spatial structure of the initial perturbation. We here consider a simple density perturbation $\delta n(r,z)$, such that $\delta f(r,z,v,t=0)=\delta n(r,z)f_0(v)$, and write $\delta n(r,z)$ as a superposition of eigenmodes
\begin{equation}\label{eq:delta_n(t=0)}
    \delta n(r,z)=\sum_{n,m}\dfrac{\delta n_{n,m}}{n_0}\Psi_{n,m}(r,z),
\end{equation}
where
\begin{equation}\label{eq:delta_n}
    \delta n_{n,m}=\dfrac{\int_{0}^{2L_{p}}dz\int_{0}^{R_{w}}dr\, r\, n_0(r) \Psi_{n,m}(r,z) \delta n(r,z) } {L_{p}\int_{0}^{R_{w}}dr\,r\dfrac{n_{0}(r)}{n_{0}}\psi_{n,m}^2(r)}.
\end{equation}
Note that the $n_0(r)$ dependence on $\delta f$ is not needed, since it is properly taken into account by writing $F(r,z,v,t)=n_0(r) f(r,v,z,t)$ as introduced in Sec.\ref{sect:wavedisp}.

This initial perturbation gives rise to a potential of the form
\begin{equation}\label{eq:delta_phi_IVP}
    \delta\phi(r,z,t)=\sum_{n,m}A_{n,m}(t)\Psi_{n,m}(r,z).
\end{equation}
A simple relation between $\delta n_{n,m}$ and $A_{n,m}(t=0)$ can be determined simply by substituting Eqs. (\ref{eq:delta_n(t=0)}) and (\ref{eq:delta_phi_IVP}) into Poisson’s equation evaluated at time $t=0$. After some algebra, it can be easily shown that
\begin{equation}\label{eq:deltanm/n0_A(0)}
    \dfrac{\delta n_{n,m}}{n_{0}}= k_{n}^{2}\lambda_{D}^{2}(n,m)\dfrac{e}{k_B T} A_{n,m}(0).
\end{equation}

By Laplace-transforming the linearized Vlasov equation in Eq. (\ref{eq:vla_linearized}) and by using the definition of $\delta f$, Eqs. (\ref{eq:delta_n(t=0)}) and (\ref{eq:delta_phi_IVP}), we obtain the coefficient
\begin{widetext}
\begin{equation}\label{eq:delta_f_analytic}
     \delta\hat{f}_{n,m}(v,p) =  \dfrac{\dfrac{e}{m}ik_{n}\hat{A}_{n,m}(p)\dfrac{\partial f_0(v)}{\partial v}}{2(ik_{n}v+p)} + \dfrac{\dfrac{-e}{m}ik_{n}\hat{A}_{n,m}(p)\dfrac{\partial f_0(v)}{\partial v}}{2(-ik_{n}v+p)} + 
       \dfrac{f_0(v)}{2} \dfrac{\delta n_{n,m}}{ n_0} \cdot \left(\dfrac{1}{p+ik_{n}v} + \dfrac{1}{p-ik_{n}v}\right),
\end{equation}
\end{widetext}
which is defined through the relation
\begin{equation}
    \delta\hat{f}(r,z,v,p) = \sum_{n,m} \delta\hat{f}_{n,m}(v,p) \Psi_{n,m} (r,z) \, .
\end{equation}

The last two terms in Eq. (\ref{eq:delta_f_analytic}) arise from the Laplace transform of the time derivative $\partial \delta f/\partial t$. Substituting into the Laplace-transformed Eq. (\ref{eq:poi_linearized}), using the relations $f_0(-v)=f_0(v)$ and using the orthogonality of the eigenmodes yield the solution
\begin{equation}
    k_{n}^{2}D_{n,m}(k_{n},p)\hat{A}_{n,m}(p)=4\pi e \delta n_{n,m}\int_{L} dv \dfrac{f_0(v)}{p+ik_{n}v},
\end{equation}
where again the $L$ on the velocity integral indicates that the velocity integral is to be taken along the Landau contour. The velocity integral on the right hand of this equation is related to plasma dielectric function 
\begin{equation}\label{eq:fM_integral}
    k_{n}^{2}D_{n,m}(k_{n},p)-k_{n}^{2}-k_{\perp n,m}^{2}-\dfrac{1}{\lambda_{D}^{2}} = \dfrac{-p}{\lambda_{D}^{2}}\int_{L}dv\dfrac{f_0(v)}{p+ik_{n}v}.
\end{equation}

Thus, the inverse transform for the mode amplitude is given by the expression
\begin{widetext}
\begin{equation}
    A_{n,m}(t)=\int_{\epsilon-i\infty}^{\epsilon+i\infty}\dfrac{dp}{2\pi i}e^{pt}\dfrac{4\pi e\delta n_{n,m}\int_{L}dv\dfrac{f_0(v)}{p+ik_{n}v}}{k_{n}^{2}D_{n,m}(k_{n},p)} = - 4\pi e \delta n_{n,m}\int_{\epsilon-i\infty}^{\epsilon+i\infty}\dfrac{dp}{2\pi i}e^{pt}\left(1-\dfrac{k_{n}^{2}+k_{\perp n,m}^{2}+\dfrac{1}{\lambda_{D}^{2}}}{k_{n}^{2}D_{n,m}(k_{n},p)}\right)\dfrac{\lambda_{D}^{2}}{p}.
\end{equation}
\end{widetext}
There is a pole at $p=0$, but Eq. (\ref{eq:fM_integral}) implies that the coefficient of that pole vanishes at $p=0$.

For each $(n,m)$, we retain the two Landau poles, which are the least damped of the remaining poles yielding the time-asymptotic result
\begin{eqnarray}\label{eq:A(t)_asymptotic_IVP}
   A_{n,m}(t)  \simeq && 4\pi e\delta n_{n,m}\left(1+\dfrac{k_{\perp n,m}^{2}}{k_n^2}+\dfrac{1}{\lambda_{D}^{2}k_n^2}\right) \nonumber \\
    &&\dfrac{k_{n}^{2}\lambda_{D}^{2}}{2(k_{n}^{2}+k_{\perp n,m}^{2})}\cos{(\omega_{n,m}t)}e^{\gamma_{n,m}t}
\end{eqnarray}
where $p_{n,m}=\pm i\omega_{n,m}+\gamma_{n,m}$ are the Landau poles, we have assumed that $\lvert \gamma_{n,m}\rvert \ll \omega_{n,m}$, and we have noted that $(\partial D_{n,m}/\partial p_{n,m}) p_{n,m} \simeq 2\omega_{p}^{2}/\omega_{n,m}^{2} \simeq 2[(k_{n}^{2}+k_{\perp n,m}^{2})/k_{n}^{2}]$ for $\omega_{n,m}\ll \omega_{p}$. 

Result (\ref{eq:A(t)_asymptotic_IVP}) is exact for the case of a cold plasma, where
\begin{equation}
  k_{n}^{2}D_{n,m}(k,0)=(k_{n}^{2}+k_{\perp n,m}^{2})\left(1+\omega_{p}^{2}(n,m)/p^{2}\right)   
\end{equation}
For a warm plasma, Eq. (\ref{eq:A(t)_asymptotic_IVP}) is only time asymptotically valid, but a more exact solution, also valid for early time, can be obtained by numerically integrating the contour integral in Eq. (\ref{eq:fM_integral}) along the vertical line where $\Re(p)=\epsilon>0$. The dielectric must be evaluated using the plasma dispersion function in Eq. (\ref{eq:TG_dispersion_relation_rewritten}), but for the Landau contour in the velocity integral is just an integral along the real $v$-axis.  Of course, all of the parameters that enter the integral must have specific numerical values.

The most important thing about Eq. (\ref{eq:A(t)_asymptotic_IVP}) is that $A_{n,m}(t)$ is proportional to $\delta n_{n,m}$; namely, a single mode can be launched simply by choosing the spatial dependence of the initial density perturbation to match that of the mode to be launched.

%%%%%%%%%%%%%%%%%%%%%%%%%%%%%%%%%%%%%%%%%%%%%%%%%%%%
\section{Linear theory of wave launching with wall electrodes }\label{sect:Launching_theory}
%%%%%%%%%%%%%%%%%%%%%%%%%%%%%%%%%%%%%%%%%%%%%%%%%%%%

In this section, we develop a linear theory of wave launching for TGWs.  In wave launching experiments, an oscillating potential $V(t)$ is applied to an electrically isolated section of the conducting wall. Suppose that in the real plasma the launching electrode extends from $z=L_{p}-\Delta z/2$ to $z=L_{p}$; then in the mirror plasma we must impose a mirror electrode that extends also from $z=L_{p}$ to $z=L_{p}+\Delta z/2$. Therefore, in the extended domain of the model, $z=0$ to $z=2L_{p}$, a single continuous electrode extends from $z=L_{p}-\Delta z/2$ to $z=L_{p}+\Delta z/2$, as shown in Fig. \ref{fig:picnum}. The potential applied to the wall changes the boundary condition on the wall to the requirement that $\delta\phi(R_{w},z,t)=V(t)$ for $L_{p}-\Delta z/2 < z < L_{p}+\Delta z/2$ and $\phi(R_{w},z,t)=0$ elsewhere. 

Let $\delta\phi^{(v)}(r,z,t)$ be the vacuum potential response to the applied voltage in the absence of a plasma. This potential satisfies Laplace’s equation
\begin{equation}
    \left(\dfrac{1}{r}\dfrac{\partial}{\partial r}r\dfrac{\partial}{\partial r}+\dfrac{\partial^{2}}{\partial z^{2}}\right)\delta\phi^{(v)}(r,z,t)=0,
\end{equation}
which subject to the above boundary condition on the wall together with the conditions that $\partial\delta\phi^{(v)}/\partial r$ vanish at $r=0$ and $\delta\phi^{(v)}(r,z,t)$ is periodic of period $2L_{p}$. The general solution can be written in the form
\begin{eqnarray}
    \delta\phi^{(v)}(r,z,t)= &&\sum_{n}V_{n}(t)\cos(k_{n}z)\dfrac{I_{0}(k_{n}r)}{I_{0}(k_{n}R_{w})}+ \nonumber \\
    &&\sum_{n}U_{n}(t)\sin(k_{n}z)\dfrac{I_{0}(k_{n}r)}{I_{0}(k_{n}R_{w})},
\end{eqnarray}
where
\begin{small}
\begin{eqnarray}
&& V_{n}(t)=\dfrac{V(t)\int_{L_{p}-\Delta z/2}^{L_{p}+\Delta z/2}dz\,\cos(k_{n}z)}{L_{p}}=\dfrac{V(t)(-1)^{n}2\sin(k_{n}\Delta z/2)}{n\pi} \label{eq:Vn(t)}, \\
&& U_{n}(t)=\dfrac{V(t)\int_{L_{p}-\Delta z/2}^{L_{p}+\Delta z/2}dz\,\sin(k_{n}z)}{L_{p}}=0. \label{eq:Un(t)}
\end{eqnarray}
\end{small}
Note that the boundary condition $\partial \phi_{v}(r,z,t)/\partial z=0$ at $z=0$ and $z=L_{p}$ is satisfied automatically because the electrode is symmetric about $z=L_{p}$.

When the plasma is present, the potential can be decomposed as $\delta\phi(r,z,t)=\delta\phi^{(v)}(r,z,t)+\delta\phi^{(p)}(r,z,t)$ where the plasma potential $\delta\phi^{(p)}(r,z,t)$ satisfies Poisson’s equation and vanishes on the conducting wall. The combination of $\delta\phi^{(p)}(r,z,t)+\delta\phi^{(v)}(r,z,t)$ then satisfies Poisson’s equation and the nonzero boundary condition on the wall. Because $\delta\phi^{(p)}(r,z,t)$ satisfies the same boundary conditions as in Secs. \ref{sect:intro}-\ref{sect:launching} , we continue to expand in the superposition of the same eigenmodes
\begin{equation}\label{eq:delta_phi_DRIV}
    \delta\phi^{(p)}(r,z,t)=\sum_{n,m}A_{n,m}^{(p)}(t)\Psi_{n,m}(r,z).
\end{equation}

Substituting the expansions
\begin{eqnarray}
  \delta\phi(r,z,t) = &&\sum_{n,m}A_{n,m}^{(p)}(t)\Psi_{n,m}(r,z)+ \nonumber \\
  &&\sum_{n}V_{n}(t)\cos(k_{n}z)\dfrac{I_{0}(k_{n}r)}{I_{0}(k_{n}R_{w})}
\end{eqnarray}
into Eq. (\ref{eq:vla_linearized}) and Laplace transforming in time  yield the following expression for the coefficient $\delta \hat{f}_{n,m}(v,p)$
\begin{widetext}
\begin{eqnarray}
    \delta\hat{f}_{n,m}(v,p) = &&\dfrac{\dfrac{e}{m}ik_{n}\hat{A}_{n,m}^{(p)}(p)\dfrac{\partial f_0(v)}{\partial v}}{2(ik_{n}v+p)} + \dfrac{\dfrac{-e}{m}ik_{n}\hat{A}_{n,m}^{(p)}(p)\dfrac{\partial f_0(v)}{\partial v}}{2(-ik_{n}v+p)} + \nonumber \\
     &&\dfrac{\dfrac{e}{m}ik_{n}\hat{V}_n(p) \dfrac{\partial f_0(v)}{\partial v}}{2(ik_{n}v+p)}\dfrac{\int_0^{R_w}dr \, r  \dfrac{n_0(r)}{n_0} \dfrac{I_{0}(k_{n}r)}{I_{0}(k_{n}R_{w})} \psi_{n,m}(r)}{\int_{0}^{R_{w}}dr\,r\dfrac{n_{0}(r)}{n_{0}}\psi_{n,m}^2(r)} + 
     \dfrac{\dfrac{-e}{m}ik_{n}\hat{V}_n(p)\dfrac{\partial f_0(v)}{\partial v}}{2(-ik_{n}v+p)}\dfrac{\int_0^{R_w}dr \, r  \dfrac{n_0(r)}{n_0} \dfrac{I_{0}(k_{n}r)}{I_{0}(k_{n}R_{w})} \psi_{n,m}(r)}{\int_{0}^{R_{w}}dr\,r\dfrac{n_{0}(r)}{n_{0}}\psi_{n,m}^2(r)} \label{eq:delta_f_num}
\end{eqnarray}
\end{widetext}

Here, there is no term such as the last bracket in Eq. (\ref{eq:delta_f_analytic}), because the initial perturbations are all zero. Taking the Laplace transform of Eq. (\ref{eq:poi_linearized}), substituting Eq. (\ref{eq:delta_f_num}), using the relation $f_0(-v)=f_0(v)$, the form for $\hat{V}_n(p)$ derived from Eq. (\ref{eq:Vn(t)}), and using the orthogonality of the eigenfunctions yields the expression
\begin{equation}
k_{n}^{2}D_{n,m}(k_{n},p)\hat{A}_{n,m}^{(p)}=C_{n,m}\hat{V}(p)K^2(k_{n},p) \, ,
\end{equation}
being
\begin{small}
\begin{equation}
    C_{n,m}=\dfrac{(-1)^{n}2\sin(k_{n}\Delta z/2)}{n\pi}\dfrac{\int_{0}^{R_{w}}dr\,r\dfrac{n_{0}(r)}{n_{0}}\psi_{n,m}(r)\dfrac{I_{0}(k_{n}r)}{I_{0}(k_{n}R_{w})}}{\int_{0}^{R_{w}}dr\,r\dfrac{n_{0}(r)}{n_{0}}\psi_{n,m}^2(r)} \, ,
\end{equation}
\end{small}
whose solution is
\begin{equation}
    \hat{A}_{n,m}^{(p)}(p)=C_{n,m}\hat{V}(p)\dfrac{K^2(k_{n},p)}{k_{n}^{2}D_{n,m}(k_{n},p)}
\end{equation}

Note that the quantity $V(t)C_{n,m}$ is simply the projection of $\delta\phi^{(v)}(r,z,t)$ onto the plasma eigenmode $\Psi_{n,m}(r,z)$, since
\begin{small}
\begin{equation}
\dfrac{\int_{0}^{2L_{p}}dz\int_{0}^{R_{w}}dr\,r\dfrac{n_{0}(r)}{n_{0}}\Psi_{n,m}(r,z)\delta\phi^{(v)}(r,z,t)}{\int_{0}^{2L_{p}}dz\int_{0}^{R_{w}}dr\,r\dfrac{n_{0}(r)}{n_{0}}\Psi_{n,m}(r,z)\Psi_{n,m}(r,z)} = V(t)C_{n,m}.
\end{equation}
\end{small}
We are interested in the projection of the full potential, $\delta\phi(r,z,t)=\delta\phi^{(v)}(r,z,t)+\delta\phi^{(p)}(r,z,t)$, onto the $(n,m)$ mode, the Laplace transform of which is given by the expression
\begin{eqnarray}
\hat{A}_{n,m}(p) &&=\hat{V}(p)C_{n,m}+\hat{A}_{n,m}^{(p)}(p) \nonumber \\
%&&= \hat{V}(p)C_{n,m}+\hat{V}(p)C_{n,m}\left(\dfrac{k_{n}^{2}+k_{\perp n,m}^{2}}{k_{n}^{2}}\dfrac{1}{D_{n,m}(k_{n},p)}-1\right) \nonumber  \\
&&= C_{n,m}\hat{V}(p)\left(\dfrac{k_{n}^{2}+k_{\perp n,m}^{2}}{k_{n}^{2}}\dfrac{1}{D_{n,m}(k_{n},p)}\right) \label{eq:Anm(p)}
\end{eqnarray}
where use has been made of Eq. (\ref{eq:dispersion relation}) to write $K^{2}(k_{n},p)$ in terms of $k_{n}^{2}D_{m,m}$. Since the $p$ dependence of $\hat{A}_{n,m}(p)$ is given by the product of two factors that depend on $p$, namely $\hat{V}(p)$ and $1/D_{n,m}(k_{n},p)$, the inverse transform can be evaluated by using the Faltung theorem \citep{morse1953methods}
\begin{equation}\label{eq:Anm(t)}
    A_{n,m}(t)=C_{n,m}\int_{0}^{t}d\tau V(\tau)g_{n,m}(t-\tau),
\end{equation}
where
\begin{eqnarray}
    g_{n,m} &&=\dfrac{k_{n}^{2}+k_{\perp n,m}^{2}}{k_{n}^{2}}\int_{\epsilon-i\infty}^{\epsilon+i\infty}\dfrac{dp}{2\pi i}e^{pt}\dfrac{1}{D_{n,m}(k_{n},p)} \nonumber \\
         &&\simeq -\omega_{n,m}\sin(\omega_{n,m}t)e^{\gamma_{n,m}t} \label{eq:gnm(t)}.
\end{eqnarray}
Here, $p_{n,m}=\pm i\omega_{n,m}+\gamma_{n,m}$ are the Landau poles, it has been assumed that $\lvert \gamma_{n,m}\rvert \ll \omega_{n,m}$ and we have noted that $\partial D_{n,m}/\partial p_{n,m} \simeq -2\omega_{p}^{2}/\omega_{n,m}^{3} \simeq -2/\omega_{n,m}(k_{n}^{2}+k_{\perp n,m}^{2})/k_{n}^{2}$ for $\omega_{n,m}\ll\omega_{p}$. Again, the last term is exact for a cold plasma, but only time asymptotic for a warm plasma. More accurate results for the last term can be obtained by numerically integrating the contour integral in Eq. (\ref{eq:gnm(t)}).

\begin{figure}
    \includegraphics[width=0.8\columnwidth]{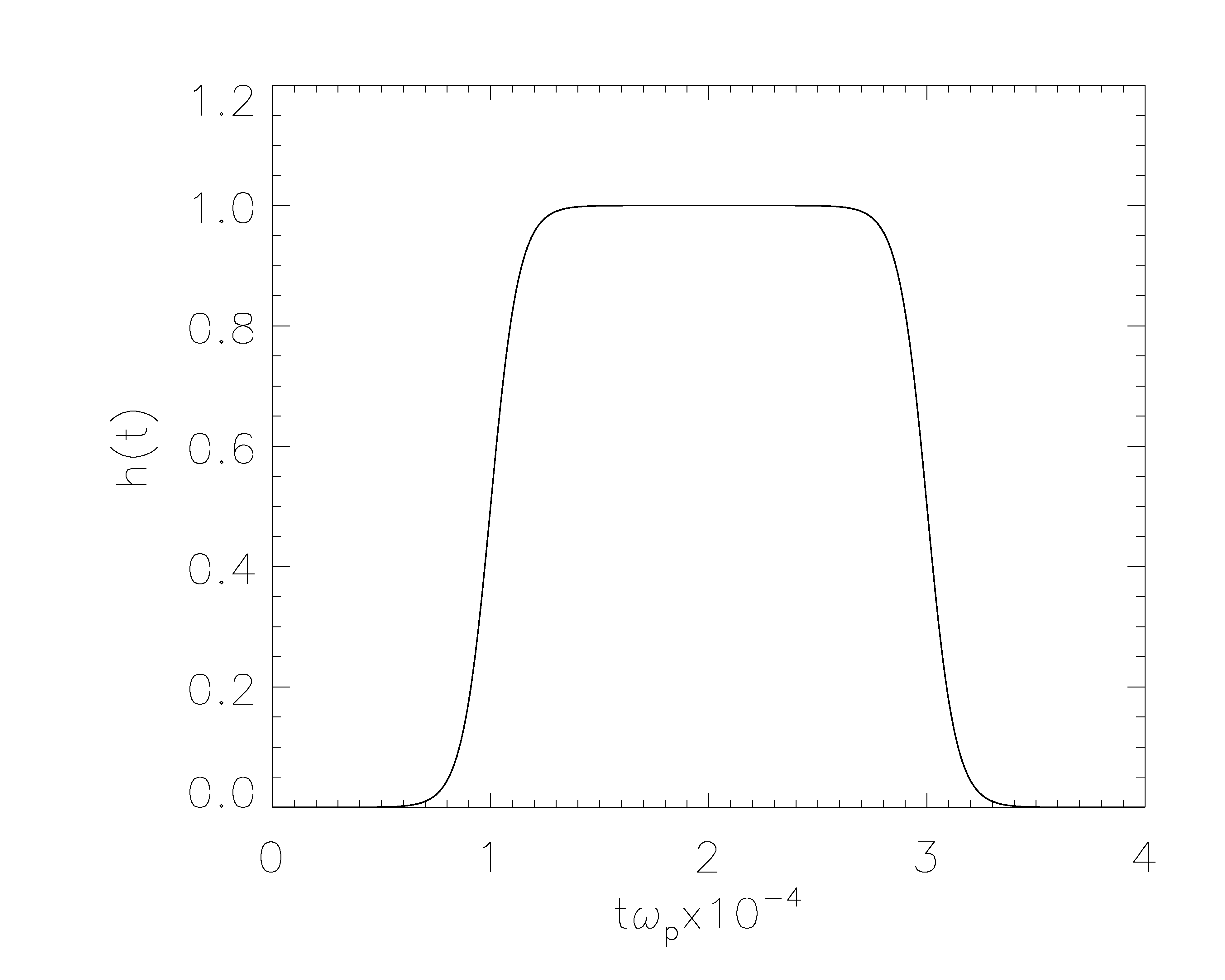}
    \caption{Time dependence of $h(t)$, for the paramaters $t_1=10000$, $t_2=30000$, $\Delta t=1300$.}                            % FIGURE 3
    \label{fig:driver_h(t)}
\end{figure}

The launching potential $V(t)$ is typically of the form
\begin{equation}\label{eq:launching_potential_V(t)}
    V(t)=V_{D}h(t)\sin(\omega_{D}t),
\end{equation}
where $\omega_{D}$ is the driver frequency, $V_{D}$ is the driver amplitude, and $h(t)$ is a function that turns on and off the driver adiabatically, thus avoiding an abrupt turn on and off of the driver field \citep{pezzi2015nonlinear,trivedi2016chirp}. An example that is analytically tractable is the function
\begin{equation}\label{eq:h(t)}
    h(t)=\dfrac{1}{2}\left[\tanh{\left(\dfrac{t-t_{1}}{\Delta t}\right)-\tanh{\left(\dfrac{t-t_{2}}{\Delta t}\right)}}\right]
\end{equation}
which is plotted in Fig. \ref{fig:driver_h(t)} for the values $t_1=10000\omega_p^{-1}$, $t_2=30000\omega_p^{-1}$, $\Delta t=1300\omega_p^{-1}$
Let us evaluate the integral in Eq. (\ref{eq:Anm(t)}) for the case of a cold plasma, where solution (\ref{eq:gnm(t)}) is exact and $\gamma_{n,m}=0$. We say that a wave is launched if it rings well after the drive voltage is turned off, and we will see that this is the case only when $(\lvert \omega_{D}-\omega_{n,m}\rvert \Delta t)\lesssim 1$. Substituting Eqs. (\ref{eq:gnm(t)})- (\ref{eq:h(t)}) into 
Eq. (\ref{eq:Anm(t)}) yields the integral
\begin{widetext}
\begin{eqnarray}
    A_{n,m}(t)&&=-\omega_{n,m}C_{n,m}V_{D}\int_{0}^{t}d\tau \ h(\tau) \sin(\omega_{D}\tau)\sin[\omega_{n,m}(t-\tau)] = \nonumber \\
    &&-\omega_{n,m}C_{n,m}V_{D}\int_{0}^{t}d\tau \ h(\tau) \dfrac{1}{2}\{\cos[(\omega_{D}+\omega_{n,m})\tau-\omega_{n,m}t]-\cos[(\omega_{D}-\omega_{n,m})\tau+\omega_{n,m}t]\}. \label{eq:Anm(t)_forcing}
\end{eqnarray}
\end{widetext}

For a time well {\it after} the voltage drive has become exponentially small, the upper limit of the integral can be extended to plus infinity. Likewise, when $t_{1}$ is large enough that the drive is exponentially small at $t=0$, the lower limit on the integral can be extended to minus infinity. When $\lvert \omega_{D}+\omega_{n,m}\rvert \Delta t  > 1$, the first cosine function in the last bracket can be neglected compared to the second, and the integral reduces to the form
\begin{small}
\begin{eqnarray}
    A_{n,m}(t)&&\simeq\Re\bigg\{\omega_{n,m}C_{n,m}V_{D}\int_{-\infty}^{+\infty}d\tau\ h(\tau) \dfrac{1}{2}e^{[i(\omega_{D}-\omega_{n,m})\tau + i\omega_{n,m}t]}\bigg\} = \nonumber \\
    &&= \Re [A_{n,m}^{\rm after}e^{(i\omega_{n,m}t)}],
\end{eqnarray}
\end{small}
where
\begin{small}
\begin{equation}\label{eq:Anm_after}
    A_{n,m}^{\rm after}=\omega_{n,m}C_{n,m}V_{D}\dfrac{\pi\Delta t}{\sinh{\dfrac{\pi\Delta t(\omega_{D}-\omega_{n,m})}{2}}}\dfrac{i}{2}[1-e^{[i(\omega_{D}-\omega_{n,m})(t_{2}-t_{1})]}]
\end{equation}
\end{small}
is the complex mode amplitude long after the drive voltage has become exponentially small. 

When the quantity $(\pi\Delta t\lvert \omega_{D}-\omega_{n,m}\rvert)/2$ is substantially larger than unity, the complex amplitude is exponentially small; that is, the mode does not ring after the drive voltage has become exponentially small.  We will see below that the mode grows adiabatically as $h(t)$ rises but then falls adiabatically when $h(t)$ decreases. Only modes for which $(\pi\Delta t\lvert \omega_{D}-\omega_{n,m}\rvert)/2$ is order unity or smaller can  be successfully launched.  This is what we mean by adiabatic frequency selection.
Another way to understand this result is to recall from Eq. (\ref{eq:Anm(p)}) that the Laplace transformed mode amplitude is given by the expression
\begin{equation}
    \hat{A}_{n,m}(p)=C_{n,m}\hat{V}(p)\left(\dfrac{k_{n}^{2}+k_{\perp n,m}^{2}}{k_{n}^{2}}\dfrac{1}{D_{n,m}(k_{n},p)}\right),
\end{equation}
so the residue at the Landau pole is proportional to $\hat{V}(p=i\omega_{n,m})$. For the launching potential discussed here, $\hat{V}(i\omega_{n,m})$ is proportional to $1/\sinh{\left(\pi\Delta t(\omega_{D}-\omega_{n,m})/2\right)}$, so the residue at the Landau pole is exponentially small if $\left(\pi\Delta t(\omega_{D}-\omega_{n,m})/2\right)$ is substantially larger than unity. Qualitatively, for adiabatic launching the frequency spectrum of $\hat{V}_{n,m}$ is so narrowly peaked that the Landau pole is outside the frequency width of the driver.

Finally, we note that in the case where $\left(\pi\Delta t(\omega_{D}-\omega_{n,m})/2\right)$ and $\lvert \omega_{D}-\omega_{n,m}\rvert(t_{2}-t_{1})$ are both small compared to unity, Eq. (\ref{eq:Anm_after}) reduces to $A_{n,m}^{\rm after}=\omega_{n,m}(t_{2}-t_{1})C_{n,m}V_{D}$.  This is the case of resonant drive, so the amplitude grows secularly for the time $t_{2}-t_{1}$.  When the quantity $\left(\pi\Delta t(\omega_{D}-\omega_{n,m})/2\right)$ is small compared to unity but the quantity $\lvert\omega_{D}-\omega_{n,m}\rvert(t_{2}-t_{1})$ is not, the factor $1-e^{[i(\omega_{D}-\omega_{n,m})(t_{2}-t_{1})]}$ in the expression for $A_{n,m}^{\rm after}$ accounts for the fact that the drive increases the amplitude of the mode for a while and then gets out of phase with the mode and begins to decrease the amplitude.

To benchmark simulations, the theoretical prediction for $A_{n,m}(t)$ in Eq. (\ref{eq:Anm(t)_forcing}) can be used to derive an expression for $\delta\phi$ to be compared to the numerical signals from simulations. Of course, in real experiments, we do not have the luxury of measuring the projection of the potential onto a particular mode, that is, measuring $A_{n,m}(t)$. Rather, electrically isolated wall sections are used to detect the presence of a mode, and the particular mode is inferred by comparing theoretical and measured frequency. On the other hand, with the simulations we can make a much more detailed comparison. Hence, we prefer to not emulate the experimental detection system.

%%%%%%%%%%%%%%%%%%%%%%%%%%%%%%%%%%%%%
\section{Numerical Model}\label{sect:nummodel}
%%%%%%%%%%%%%%%%%%%%%%%%%%%%%%%%%%%%%
For the simulations, we scale time with $\omega_{p}^{-1}$, length with $\lambda_{D}$ and potential with $k_B T/e$ defining the scaled variables
\begin{equation}
    t'=t\omega_{p}, \,\,\,\,\, z'=\dfrac{z}{\lambda_{D}},\,\,\,\,\, r'=\dfrac{r}{\lambda_{D}} \, , \,\,\,\,\, \phi'=\dfrac{e\phi}{k_B T}.
\end{equation}
On can think of this scaling as choosing the unit of time as $\omega_{p}^{-1}$, the unit of length as $\lambda_{D}$ and the unit of potential as $k_BT/e$, and these three choices determine all of the other units. Therefore, the unit of velocity is $\lambda_{D}\omega_{p}=v_{th}$, and the scaled thermal velocity is $v^\prime_{th}=1$.

\begin{figure}
\includegraphics[width=\columnwidth]{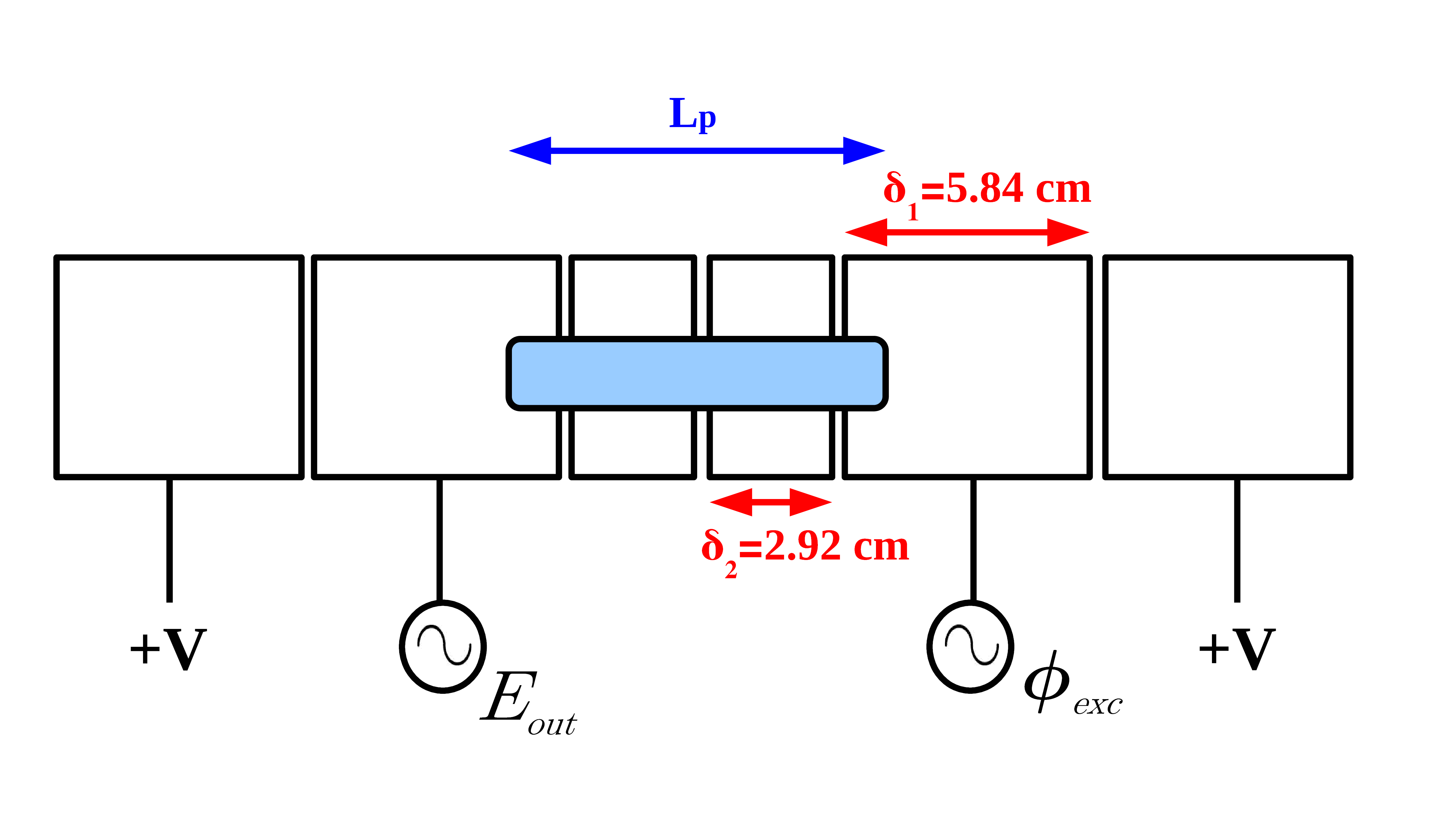}
\caption{A sketch of the experimental setup of \citet{anderegg2009electron,anderegg2009waveparticle}.} \label{fig:picexp}           % FIGURE 4
\end{figure}

In order to make close contact to the laboratory experiments described by \citet{anderegg2009electron,anderegg2009waveparticle}, we consider a non-neutral plasma column of single-ionized magnesium of density of about $2\times 10^{7}{\rm cm}^{-3}$, charge $e$ and mass $m=Am_p$, with $A=24$ the magnesium mass number. The physical dimensions of the trap are $L_{p}=9 {\rm cm}, R_{p}=0.45 {\rm cm}$, and $R_{w}=2.86, {\rm cm}$ and the scaled ones vary with $\lambda_D$, that is determined by the plasma temperature and density. In this paper, we choose $k_B T=0.01 {\rm eV}$ ($k_B T=0.5 {\rm eV}$) when simulating a cold (warm) plasma. A schematic representation of the experimental setup is shown in Fig. \ref{fig:picexp}. In these experiments, an external driving fluctuating potential $\phi_{exc}$ is usually applied to one electrode, whose length is $\delta_1=5.84$cm, to launch waves in the plasma column. The electrode is located at one longitudinal end of the the external conducting cylinder at a radial distance $R_w$ from the main axis of the plasma column.  Then, the plasma response $E_{out}$ is collected at the opposite longitudinal end. In between two electrodes, other two cylindrical sectors of length $\delta_2=2.92$cm are located, as shown in Fig. \ref{fig:picexp}.

In terms of scaled variables, the drift kinetic and Poisson’s equations take the form
\begin{eqnarray}
&&\dfrac{\partial f'}{\partial t'}+v'\dfrac{\partial f'}{\partial z'} - \dfrac{\partial \phi'}{\partial z'}\dfrac{\partial f'}{\partial v'} = 0  \label{eq:vla_scaled}, \\
&&\left(\dfrac{1}{r'}\dfrac{\partial}{\partial r'}r'\dfrac{\partial}{\partial r'} + \dfrac{\partial^{2}}{\partial z'^{2}}\right)\phi'=\dfrac{-n_{0}(r)}{n_{0}}\int dv' \,f', \label{eq:poi_scaled}
\end{eqnarray}
where $f'=v_{th}f$. 

The spatial domain for the simulations is $D_s=[0,2L_{p}']\times[0,R_{w}']$ and the velocity domain is $[-6,6]$, where we recall that $v'_{th}=1$. Outside the velocity domain, the distribution function is set equal to zero, while periodic boundary conditions are implemented in physical space. We remark that doubling the plasma column and implementing periodic boundary conditions allow for mimicking particle and wave reflections at the physical ends of the columns, that is, at $z = 0$ and $z = L_p$, as well as imposing that the wave electric field is zero there, thus being appropriate for matching on to the vacuum field beyond the end of the column \citep{prasad1983waves}. Hereafter, we assume that all variables are normalized. For the sake of simplicity, we avoid to report the prime $'$ on the variables.

The algorithm employed for the numerical solution of the kinetic equation, Eq. (\ref{eq:vla_scaled}), is based on the well-known time splitting method first proposed in 1976 by \citet{cheng1976integration}. Time splitting consists in separating the evolution of the particle distribution function in phase space into subsequent translations, first in physical space and then in velocity space. This allows to reduce the phase-space integration of the kinetic equation to the integration of two equivalent hyperbolic advection equations in physical space and velocity space respectively. A finite-difference fourth-order upwind scheme, correct up to third order in the mesh size in both physical and velocity space, is employed to evaluate spatial and velocity derivatives of the distribution function (see Refs.~\citep{mangeney2002numerical, valentini2005self,valentini2007hybrid,pezzi2013eulerian,pezzi2014kinetic} for more details on the numerical algorithm). 

For the numerical integration of the Poisson equation, Eq. (\ref{eq:poi_scaled}), we employ a standard spectral method based on the Fast Fourier Transform routine in the $z$ direction, where periodic boundary conditions are implemented. Then, we use fourth-order (correct up to third order) finite-difference schemes for both first and second derivatives of the electrostatic potential in the $r$ direction. Boundary conditions in $r$, as specified in Sec. \ref{sect:wavedisp}, are the following:
\begin{equation}
\left. \frac{\partial\phi}{\partial r}\right |_{r=0}=0; \;\;\; \phi(R_w,z,t)=\phi_{exc}(z,t) \;\; .
\label{eq:radial_BC}
\end{equation}
In this situation, integrating Eq. (\ref{eq:poi_scaled}) requires the solution of a linear, pentadiagonal system of equations, whose solution can be easily obtained by means of a standard linear algebra routine. The external driving potential $\phi_{exc}$ is modeled as
\begin{equation} \label{eq:bc_phiv_rad}
\begin{aligned}
   & \phi_{exc}(z,t)  =\begin{cases}
    V(t) &  L_{p}-\Delta z/2 <z< L_{p}+\Delta z/2, \\
    0 & \rm{elsewhere}
    \end{cases}
\end{aligned}
\end{equation}
where $\Delta z=2L_p/2-2\delta_2$, as shown in Fig.  \ref{fig:picexp}, while $V(t)$ is determined by Eq.  (\ref{eq:launching_potential_V(t)}) with $h(t)$ given by Eq. (\ref{eq:h(t)}). The values of the parameters $t_1$, $t_2$ and $\tau$ will be specified in the following.

Typical number of gridpoints employed to discretize the numerical domain are $N_z=128$ gridpoints in the periodic $z$-direction, $N_r=256$ gridpoints in the radial direction, and $N_v=1201$ in the velocity direction. The time step is always set in such a way that the Courant-Friedrichs-Lewy (CFL) condition for the numerical stability of the algorithm is satisfied~\citep{peyret1983computational}.

Since the total length of the numerical box along the $z$-direction is twice the length of the plasma column, our analysis will be restricted to the left half of the total numerical box. Then, we solved numerically Eqs. (\ref{eq:vla_scaled}) and (\ref{eq:poi_scaled}) with the correct radial boundary conditions and periodic boundary condition in $z$, to reproduce the excitation of TGWs and EAWs, as in a real experiment with non-neutral plasmas.

%%%%%%%%%%%%%%%%%%%%%%%%%%%%%%%%%%%%%
\section{Numerical simulations}\label{sect:numsim}
%%%%%%%%%%%%%%%%%%%%%%%%%%%%%%%%%%%%%
In this section, we first discuss accurate tests of the numerical code performed by simulating (i) the excitation of linear TG modes as an initial value problem and (ii) the process of launching TG waves in the trap through an external driver; in both cases, we compare the results obtained through the simulations to the theoretical predictions discussed in Secs. \ref{sect:intro}-\ref{sect:Launching_theory}. Then, we present and discuss the numerical results of the excitation of TGWs and EAWs through an external driving potential in conditions of the experiments by \citet{anderegg2009electron,anderegg2009waveparticle}. To make contact with the theory, numerical simulations in Secs. \ref{ivptest} and \ref{drivtest} have been run in the case of a cold plasma, whose equilibrium temperature is $k_B T=0.01 {\rm eV}$, while equilibrium temperature has been increased to $k_B T=0.5 {\rm eV}$ in the simulations discussed in Sec. \ref{labsim}, where a plasma in conditions close to real experiments has been modeled.

\begin{figure*}
    \subfloat{{\includegraphics[width=0.48\textwidth]{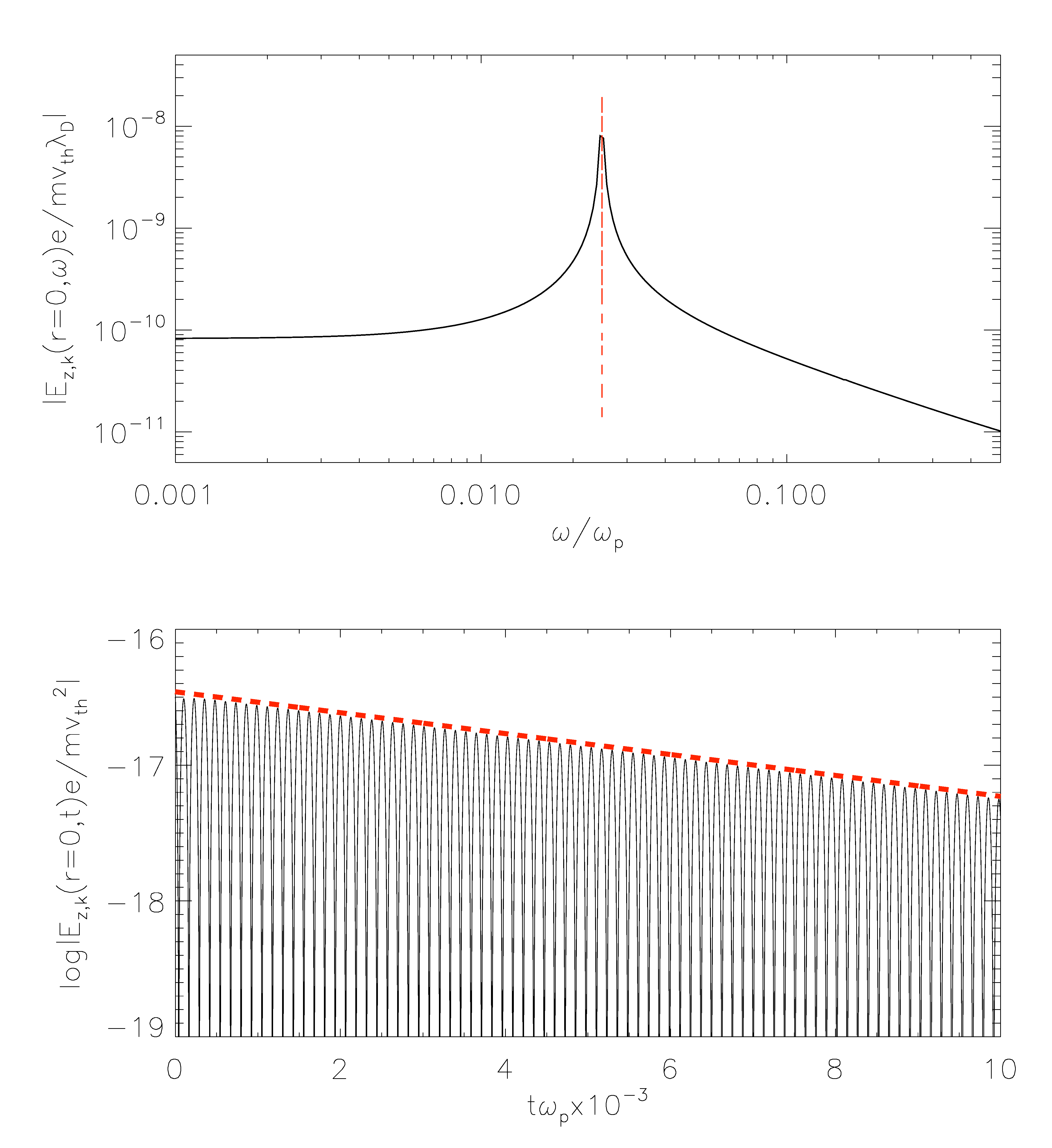} }}                                    % FIGURE 5
    \subfloat{{\includegraphics[width=0.48\textwidth]{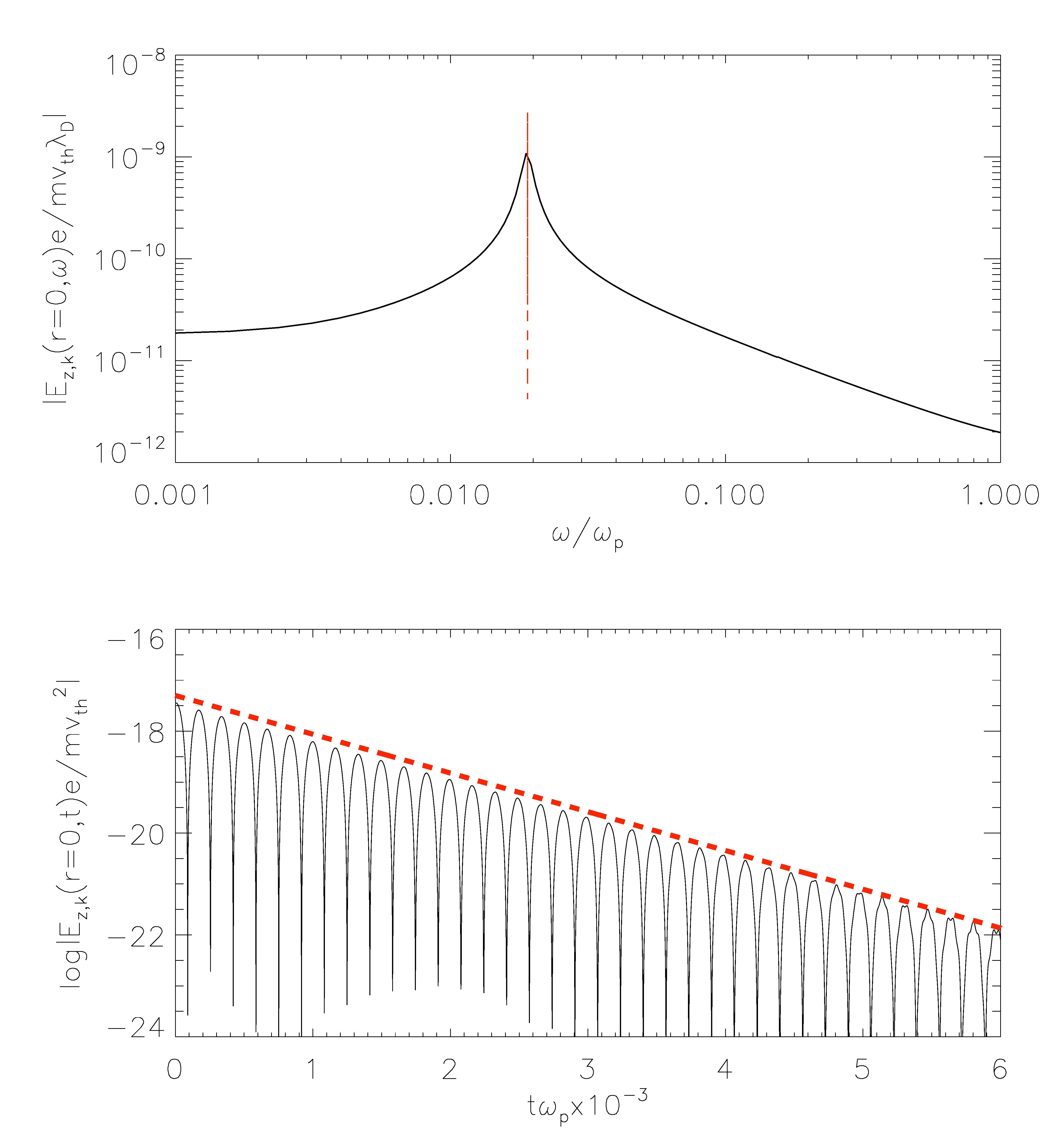} }}
    \caption{Dependence of $|E_{z,k}(r=0,\omega)|$ on $\omega$ (top row) and time evolution of $\log|E_{z,k}(r=0,t)|$ (bottom row) for initial value problem simulations in which mode $(1,2)$ (left) and mode (1,3) (right) are excited at $t=0$. The red-dashed lines in the top and in the bottom plots represent the theoretical predictions derived in section \ref{sect:launching}.}
    \label{fig:IVP_simulations}
\end{figure*}

\subsection{Excitation of TG waves from an initial perturbation}\label{ivptest}
In these simulations, the initial equilibrium is perturbed at $t=0$ through the disturbance in Eq. (\ref{eq:delta_n(t=0)}), in which we restrict the sum to a single eigenmode with specific values of $n,m$ and the amplitude of the mode is $\delta n_{n,m}=10^{-6}$. The numerical results for the $z$-component of the electric field $E_z$ are compared with theoretical predictions: in particular, from the theory in Sec. \ref{sect:launching}, we derive the theoretical prediction for the electric field perturbations as $E_z=-\partial \delta\phi/\partial z$, where $\delta\phi$ is given by Eq. (\ref{eq:delta_phi_IVP}). Specifically, we evaluate the Fourier transform along the periodic direction $z$ of $E_z(r=0,z,t)$ (taken at the radial position $r=0$ and scaled by $mv_{th}\omega_p/e$), obtaining the amplitude $E_{z,k}(r=0,t)$, and the Fourier transform both in $z$ and $t$ of the same signal obtaining the amplitude $E_{z,k}(r=0,\omega)$. 

\begin{figure}
   \includegraphics[width=0.5\textwidth]{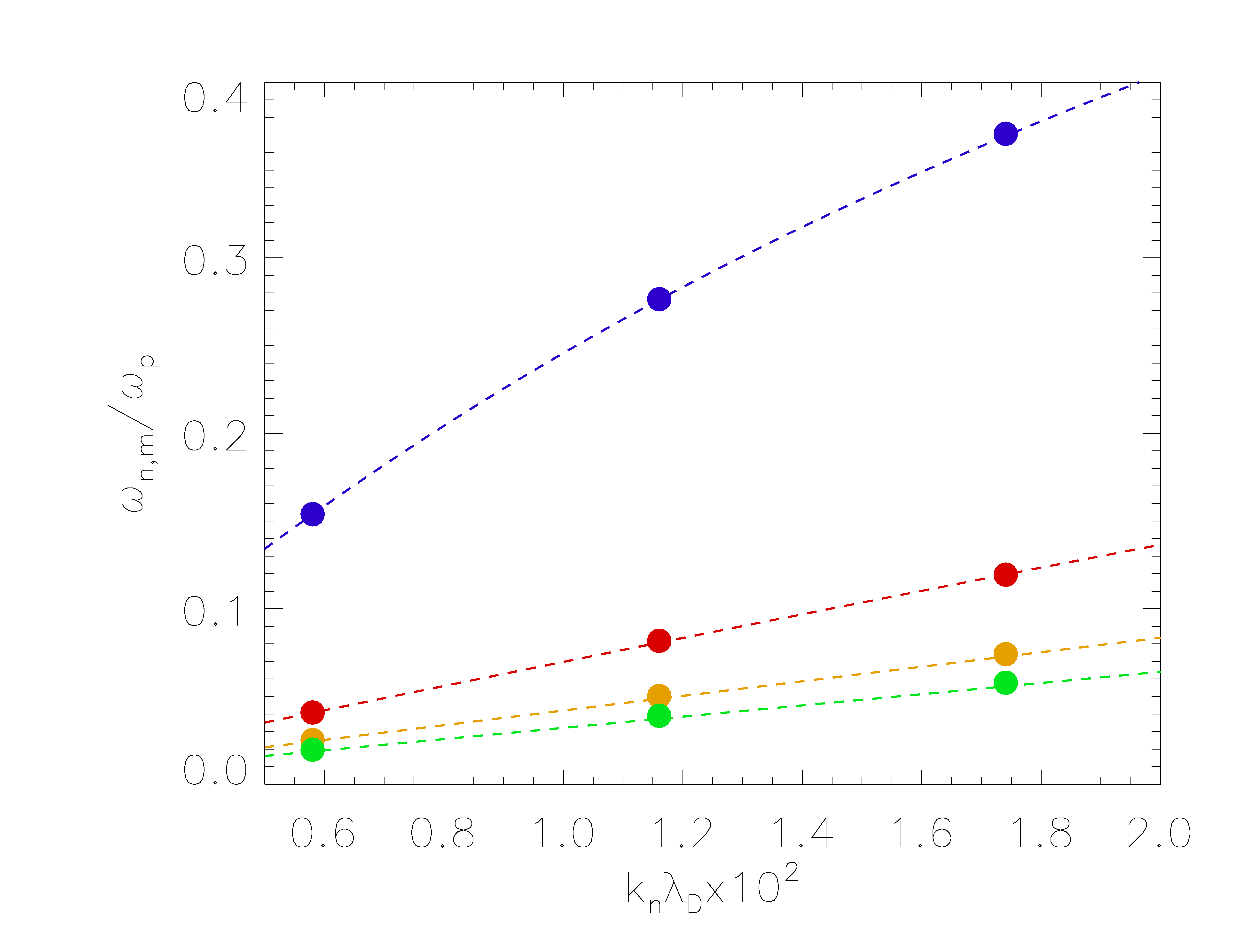}   % FIGURE 6
    \caption{$\omega_{n,m}$ as a function of $k_{n}$ for initial value problem simulations, in which eigenmodes  with  $n=1,2,3$ and $m=0,1,2,3$ have been launched at $t=0$; the dots represent the values of the real part of the mode frequency from the simulations, while the dashed lines represent the theoretical prediction in Eq. (\ref{eq:real_frequency}), for modes with $m=0$ (blue), $m=1$ (red), $m=2$ (orange) and with $m=3$ (green).}
    \label{fig:omek}
\end{figure}

In Fig. \ref{fig:IVP_simulations}, we show $|E_{z,k}(r=0,\omega)|$ from the simulation as a function of $\omega$ (top row) and the time evolution of the logarithm of $|E_{z,k}(r=0,t)|$ (bottom row); left column in the figure refers to a simulation in which the mode $(1,2)$ has been excited at $t=0$, while the right column to a simulation in which the mode $(1,3)$ has been launched as initial perturbation. The red vertical lines in the plots in the top row represent the theoretical predictions for wave frequency in Eq. (\ref{eq:real_frequency}) for mode $(1,2)$ and $(1,3)$, respectively, while the red-dashed lines in the plots in the bottom row represent the theoretical predictions for the time evolution of the logarithm of $|E_z(r=0,z,t)|$, obtained from the theory as explained above, for modes $(1,2)$ and $(1,3)$, respectively. We notice in the top plots in this figure that only the expected frequency is excited. As it is clear from the plots in Fig. \ref{fig:IVP_simulations} an excellent agreement between numerical results and theoretical predictions is found. 

\begin{figure*}
    \subfloat{{\includegraphics[width=0.48\textwidth]{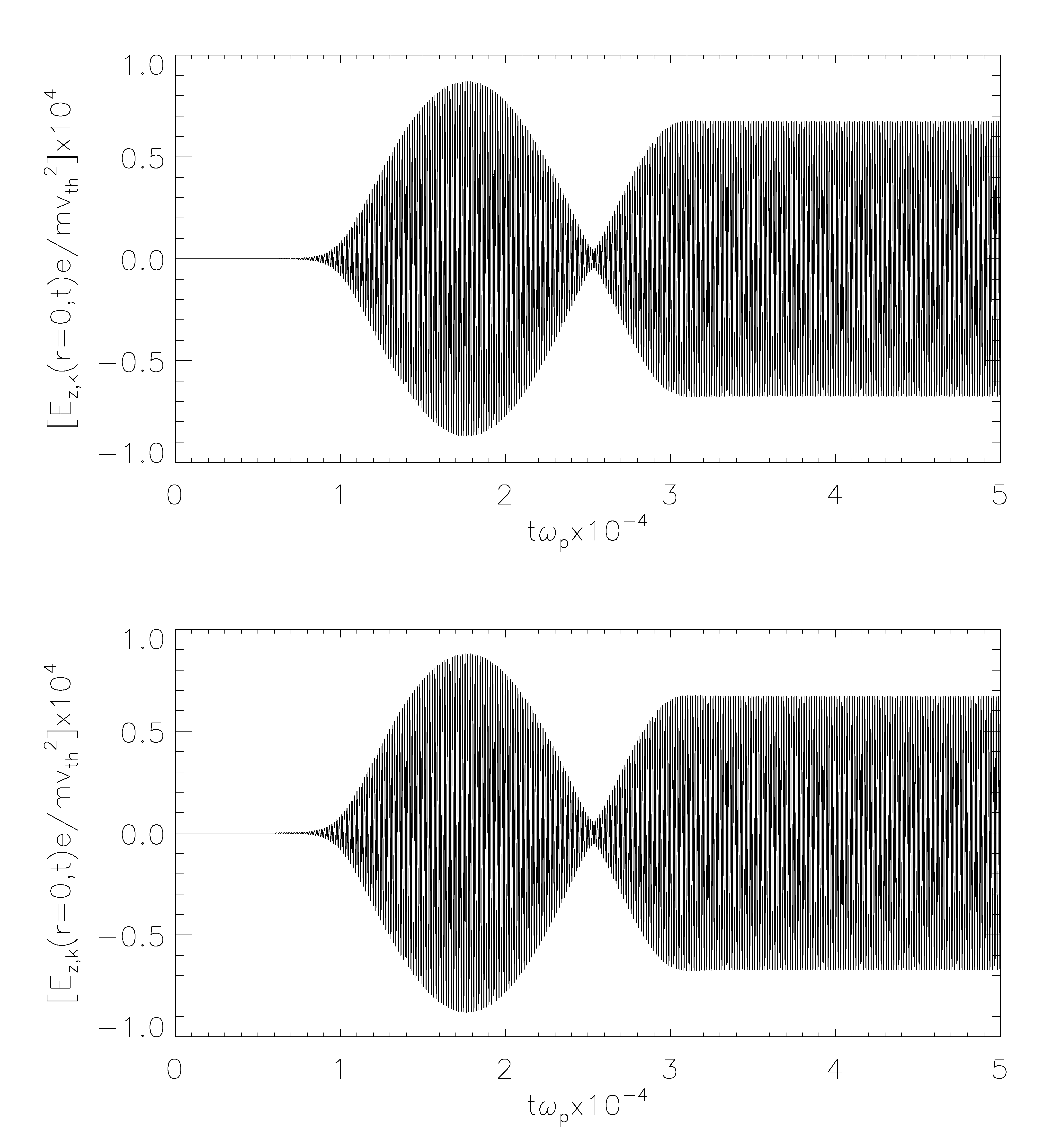} }}                              % FIGURE 7
    \subfloat{{\includegraphics[width=0.48\textwidth]{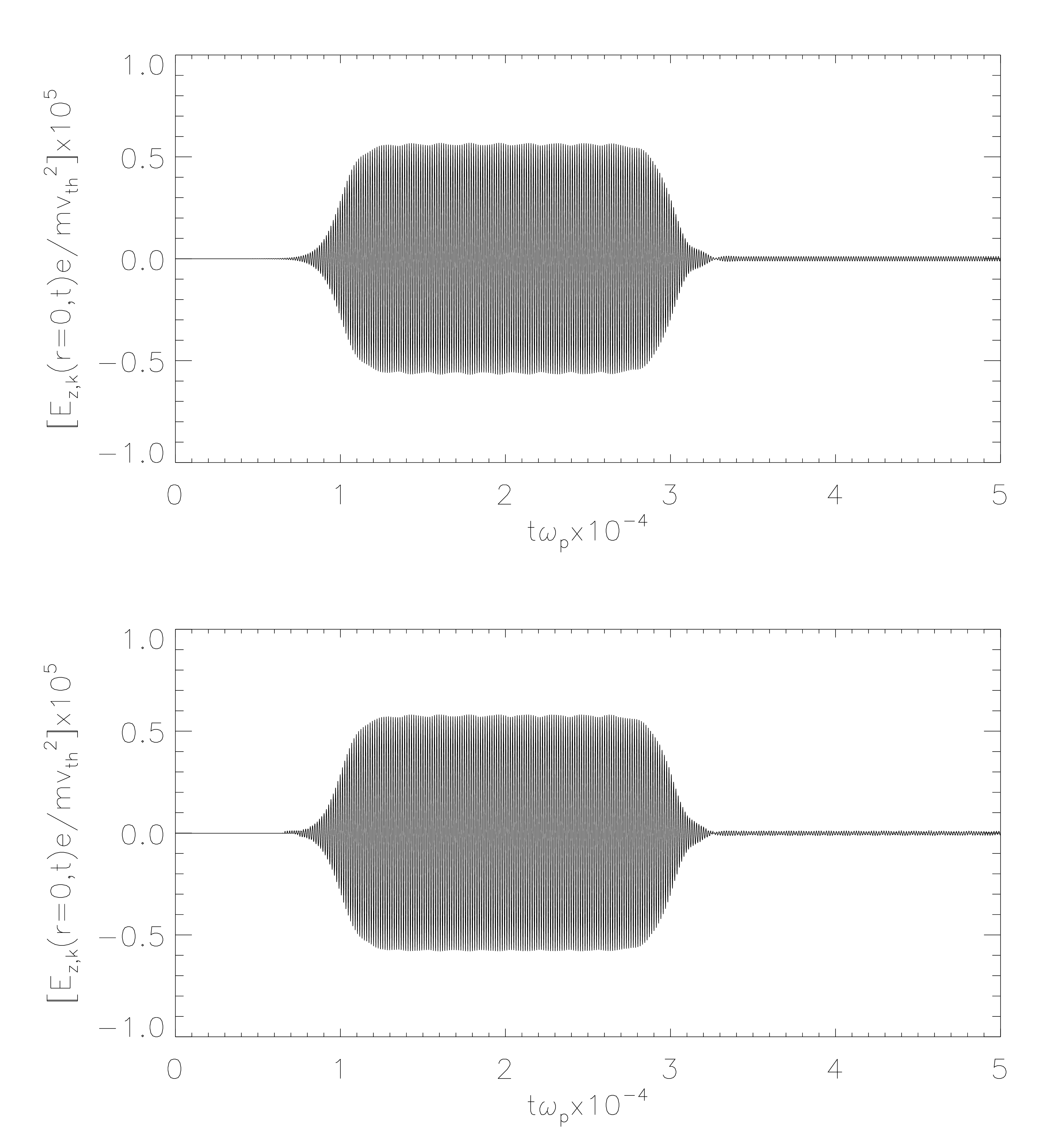}}} 
    \caption{Time evolution of $\log|E_{z,k}(r=0,t)|$ for simulations in which the plasma is driven through an external fluctuating potential with frequency $\omega_D=1.01\omega_{1,1}$ (left column, on resonance simulation) and $\omega_D=1.1\omega_{1,1}$ (right column, off resonance simulation); top row in the figure shows the theoretical prediction derived in section \ref{sect:Launching_theory}, while in the bottom row  the numerical results are reported.}
    \label{fig:DRIV_simulations}
\end{figure*}

In Fig. \ref{fig:omek} we show $\omega_{n,m}$ as a function of $k_{n}$ for twelve different initial value problem simulations, in which we considered as initial perturbations single eigenmodes with  $n=1,2,3$ and $m=0,1,2,3$: the dots here represent the values of the real part of the mode frequency from the simulations, evaluated by Fourier analyzing $E_z(r=0,z,t)$, while the dashed lines represent the theoretical curves in Eq. (\ref{eq:real_frequency}). Results for modes with $m=0$ are indicated in blue, for modes with $m=1$ in red, for modes with $m=2$ in orange and with $m=3$ in green. Also here a good agreement between numerical and theoretical results can be appreciated.

As an additional check on the numerical simulations, we controlled the value of the electric field at each end of the plasma column in time: this value remains always smaller than the maximum value of the electric field by a factor of $10^{-9}$. Moreover, we also monitored the conservation of mass, total energy, and entropy, getting percentage relative variations of $2 \times 10^{-3}$, $5 \times 10^{-3}$, and $ 3 \times 10^{-2}$, respectively.

%%%%%%%%%%%%%%%%%%%%%%%%%%%%%%%%%%%%%
\subsection{Launching TG waves through an external driver}\label{drivtest}
%%%%%%%%%%%%%%%%%%%%%%%%%%%%%%%%%%%%%
In this case, TG waves are launched in a cold plasma at equilibrium through the external driver in Eq. (\ref{eq:bc_phiv_rad}); the dependence of the function $h(t)$ is shown in Fig. \ref{fig:driver_h(t)} and we used the driver parameters $t_1=10000$, $t_2=30000$, $\Delta t=1300$, and $V_D=0.005$. The maximum time of these simulations is $t_{max}\simeq 50000$. Two different simulations with a different driver frequency $\omega_D$ were performed: one close to the theoretical frequency of the mode (1,1) (on resonance simulation with $\omega_D=1.01\omega_{1,1}$) and the other significantly larger than $\omega_{1,1}$ (off resonance simulation with $\omega_D=1.1\omega_{1,1}$). 
Also, for these simulations, we compared numerical results and theoretical predictions by focusing on the $z$-component of the electric field taken at the radial position $r=0$, $E_z(r=0,z,t)$; the theoretical prediction is obtained by evaluating $A_{n,m}$ from Eq. (\ref{eq:Anm(t)_forcing}) and using it to calculate $\delta\phi$ and, consequently, $E_z=-\partial\delta\phi/\partial z$.

The comparison of theoretical prediction and numerical results for the amplitude $E_{z,k}(r=0,t)$ is shown in Fig. \ref{fig:DRIV_simulations}, where the left (right) column refers to the on (off) resonance simulation; in the top row, we plot the theoretical prediction, while the numerical signals are shown in the bottom row. It is worth noting that, in the on resonance simulation, the electric signal survives long after the driver is turned off (at $t\simeq 35000$, as shown in Fig. \ref{fig:driver_h(t)}) ringing at a nearly constant amplitude, while no plasma response is recovered after the driver is turned off in the case of the off resonance simulation. Again, Fig. \ref{fig:DRIV_simulations} shows a very nice agreement between numerical results and analytical predictions.

%%%%%%%%%%%%%%%%%%%%%%%%%%%%%%%%%%%%%%%%%%%%%%%%%%%%%%%%%%%%%%%%%%%%%%%%%
\subsection{Simulating real laboratory experiments}\label{labsim}
%%%%%%%%%%%%%%%%%%%%%%%%%%%%%%%%%%%%%%%%%%%%%%%%%%%%%%%%%%%%%%%%%%%%%%%%%

\begin{figure*}    
\centering
  \begin{minipage}{0.24 \textwidth}  % FIGURE 8
    \includegraphics[width=\textwidth]{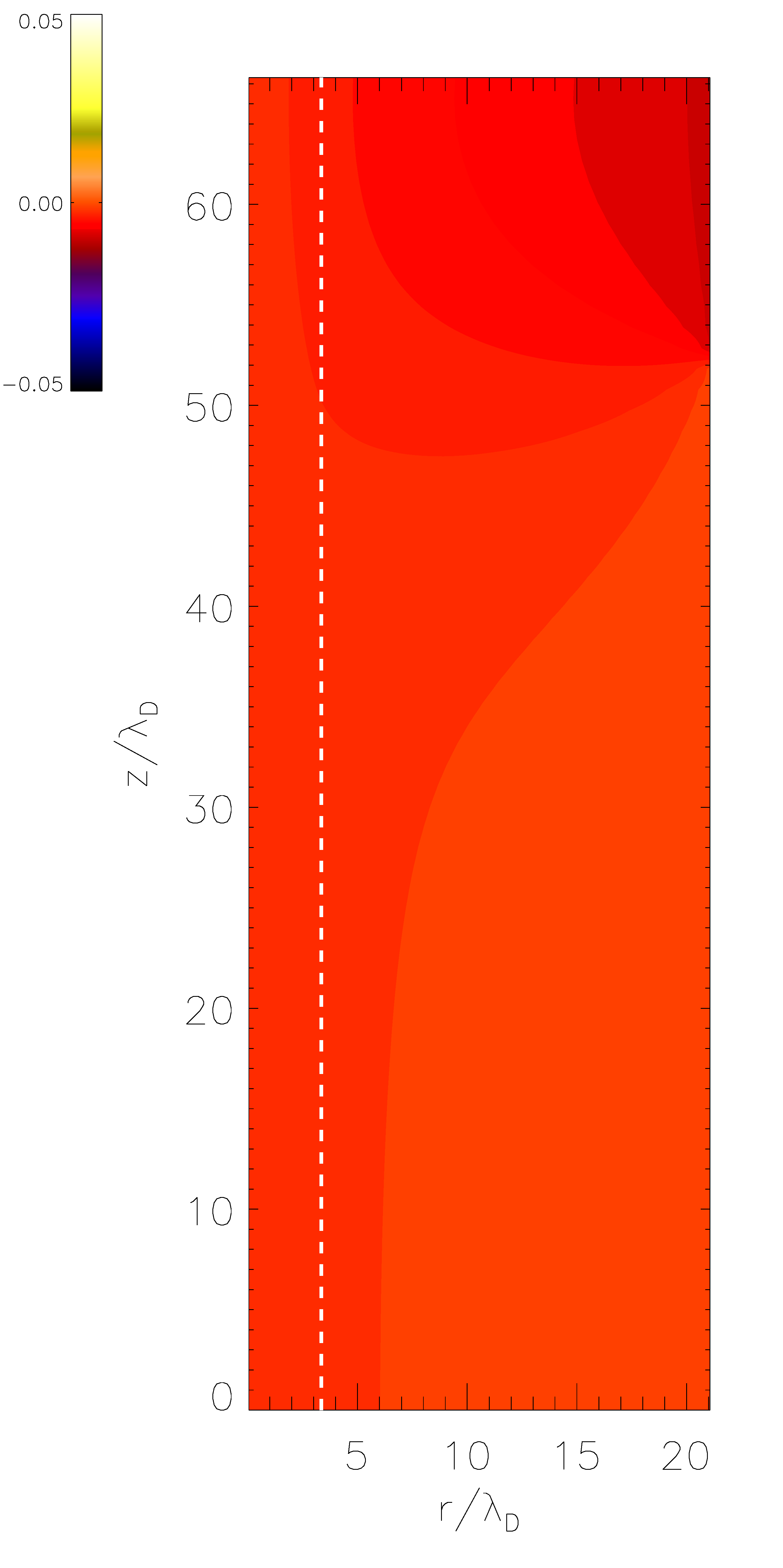}
\end{minipage}
 \hfill
  \begin{minipage}{0.24 \textwidth}
    \includegraphics[width=\textwidth]{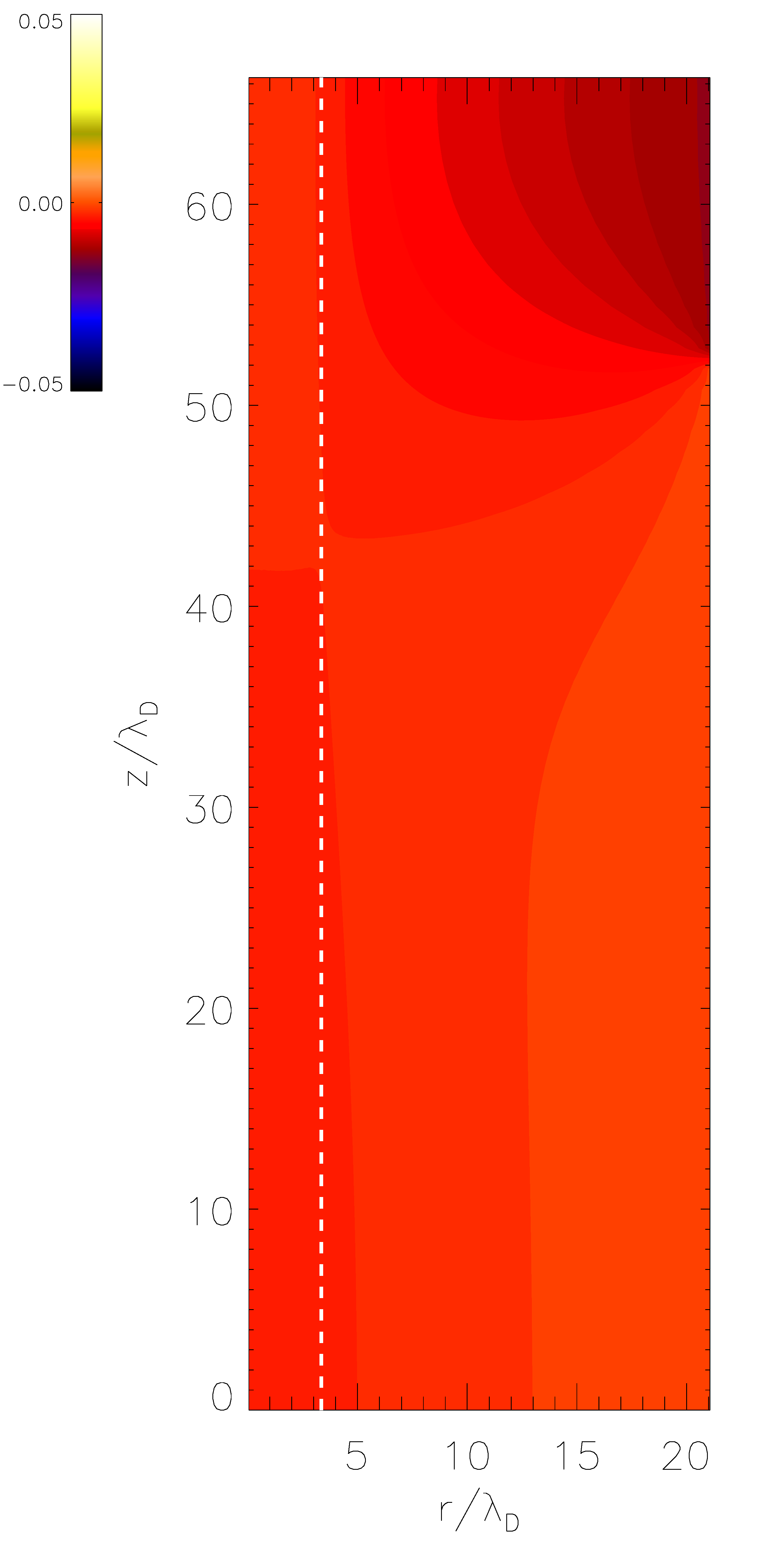}
  \end{minipage}
 \hfill
  \begin{minipage}{0.24 \textwidth}
      \includegraphics[width=\textwidth]{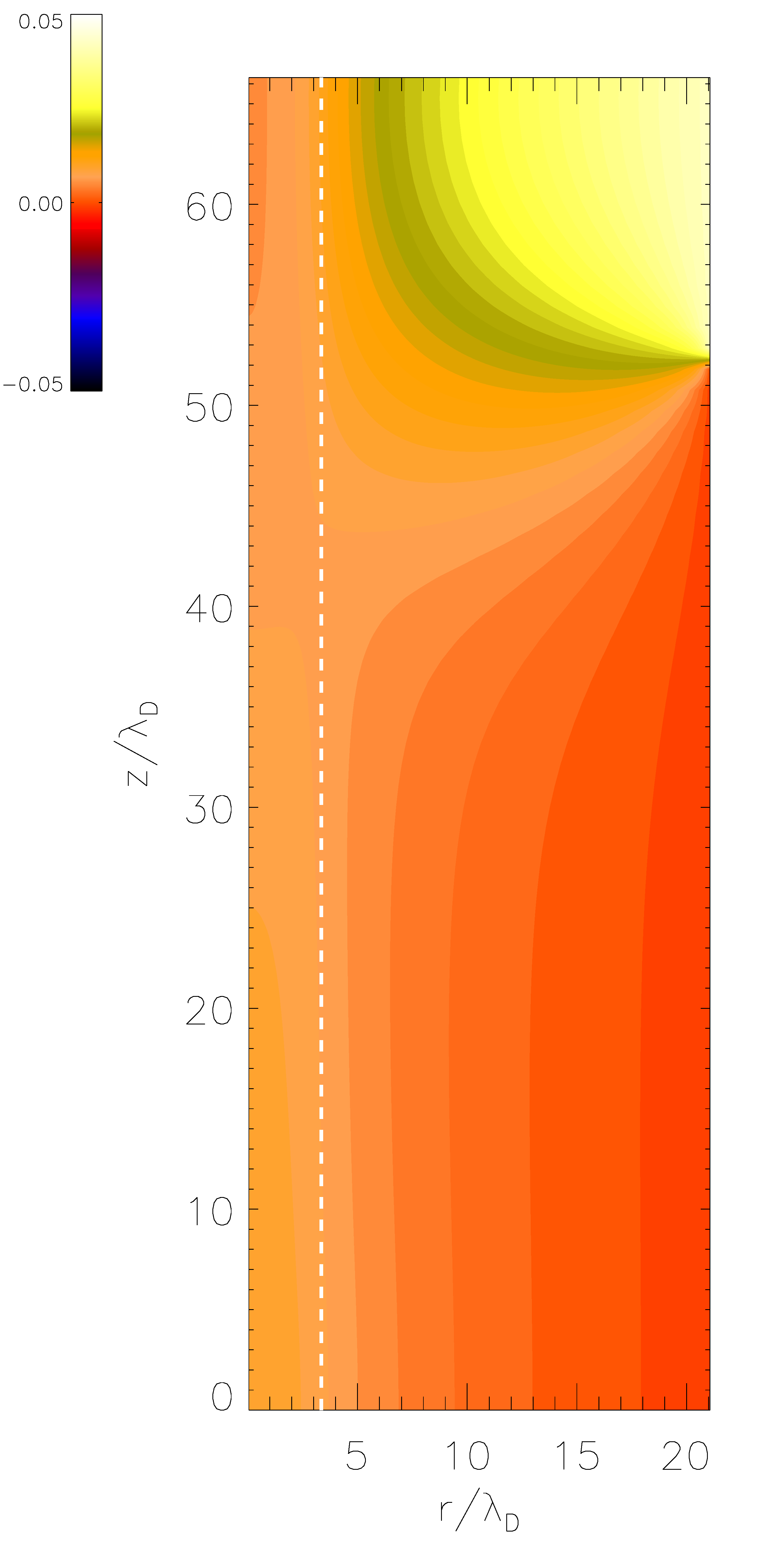}
  \end{minipage}   
 \hfill 
  \begin{minipage}{0.24 \textwidth}
      \includegraphics[width=\textwidth]{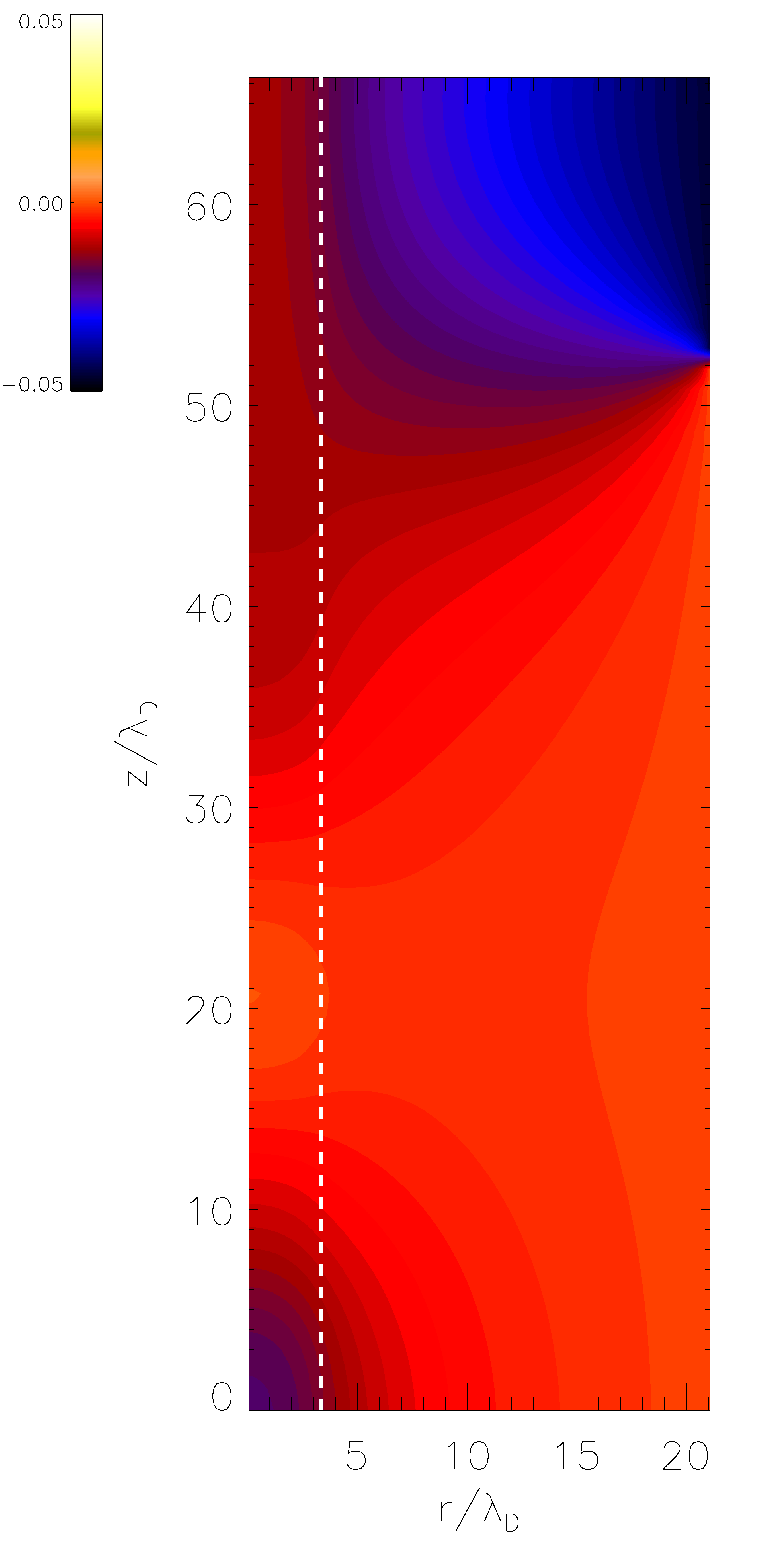}
  \end{minipage} 
 \hfill
  \begin{minipage}{0.24 \textwidth}
      \includegraphics[width=\textwidth]{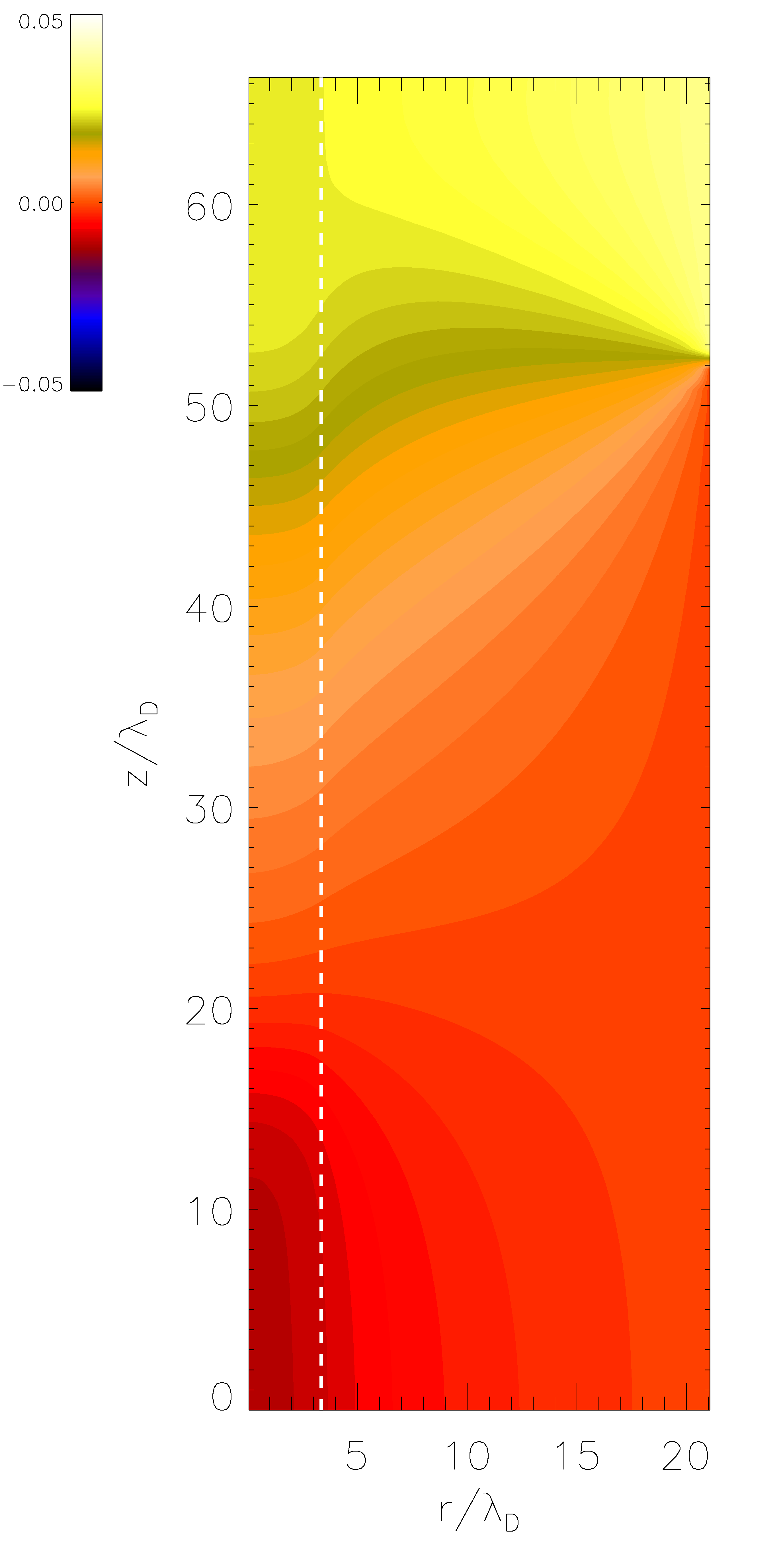}
  \end{minipage} 
 \hfill
  \begin{minipage}{0.24 \textwidth}
      \includegraphics[width=\textwidth]{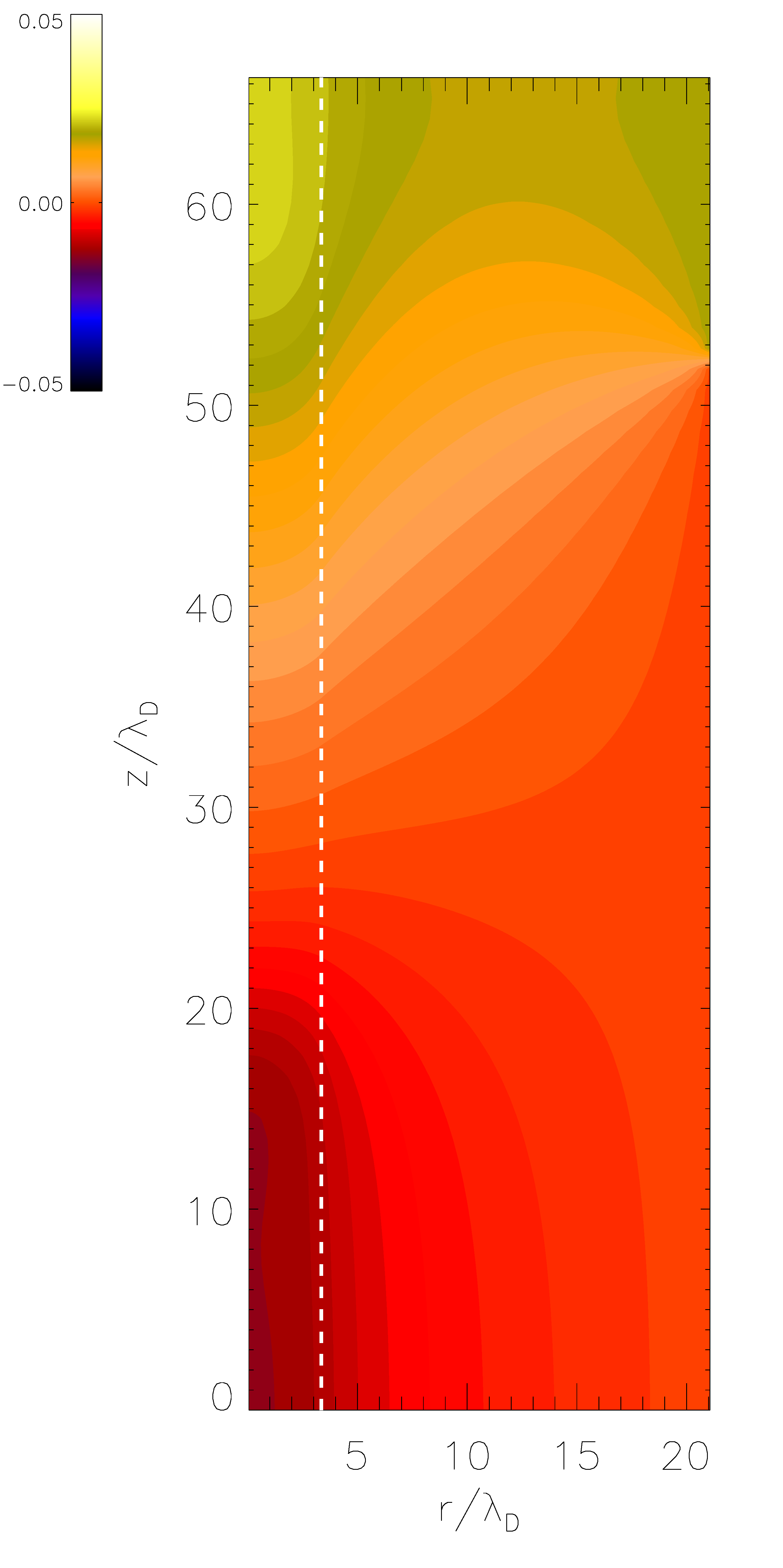}
  \end{minipage}
  \hfill
  \begin{minipage}{0.24 \textwidth}
        \includegraphics[width=\textwidth]{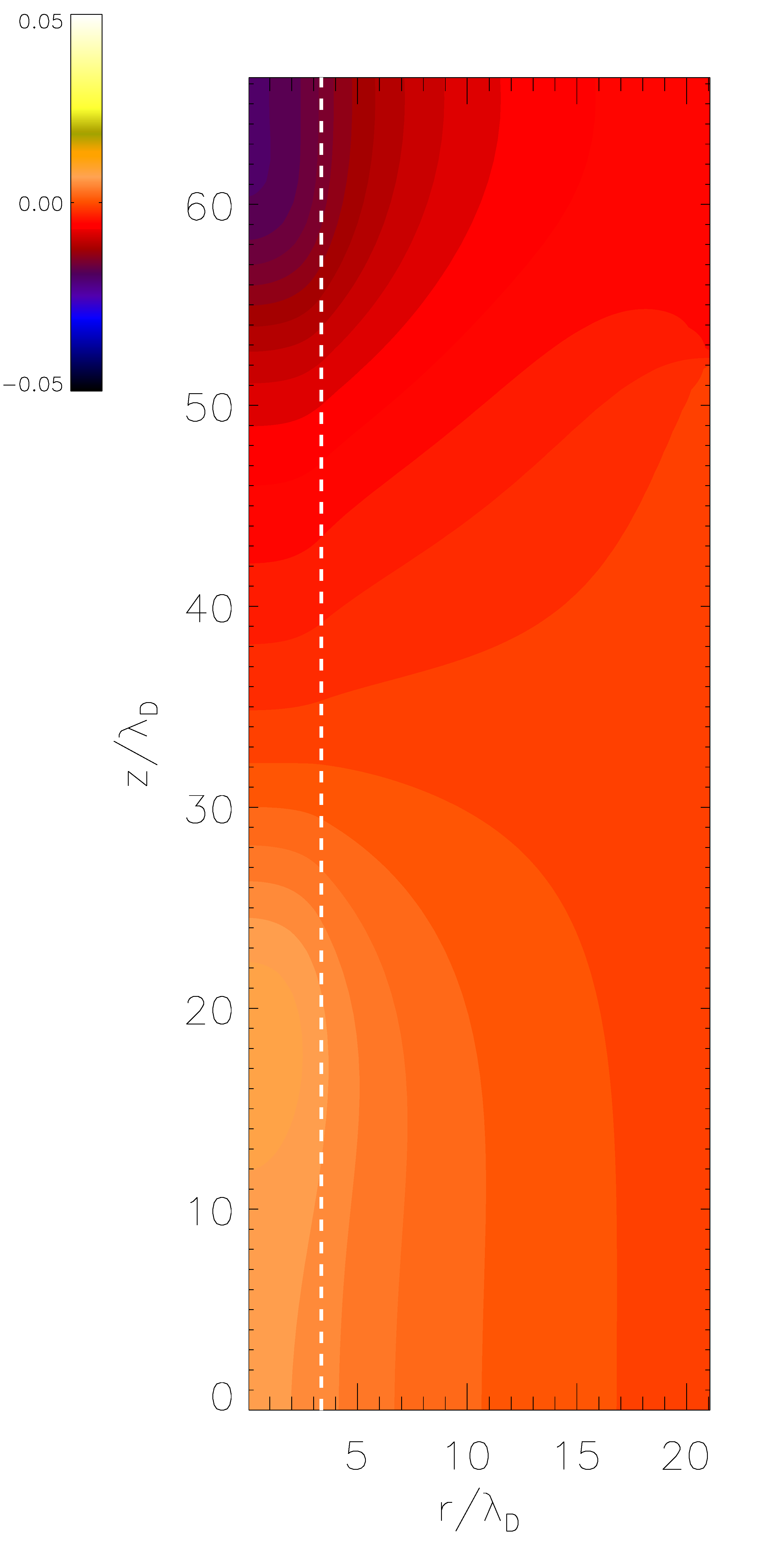}
  \end{minipage}
 \hfill
  \begin{minipage}{0.24 \textwidth}
        \includegraphics[width=\textwidth]{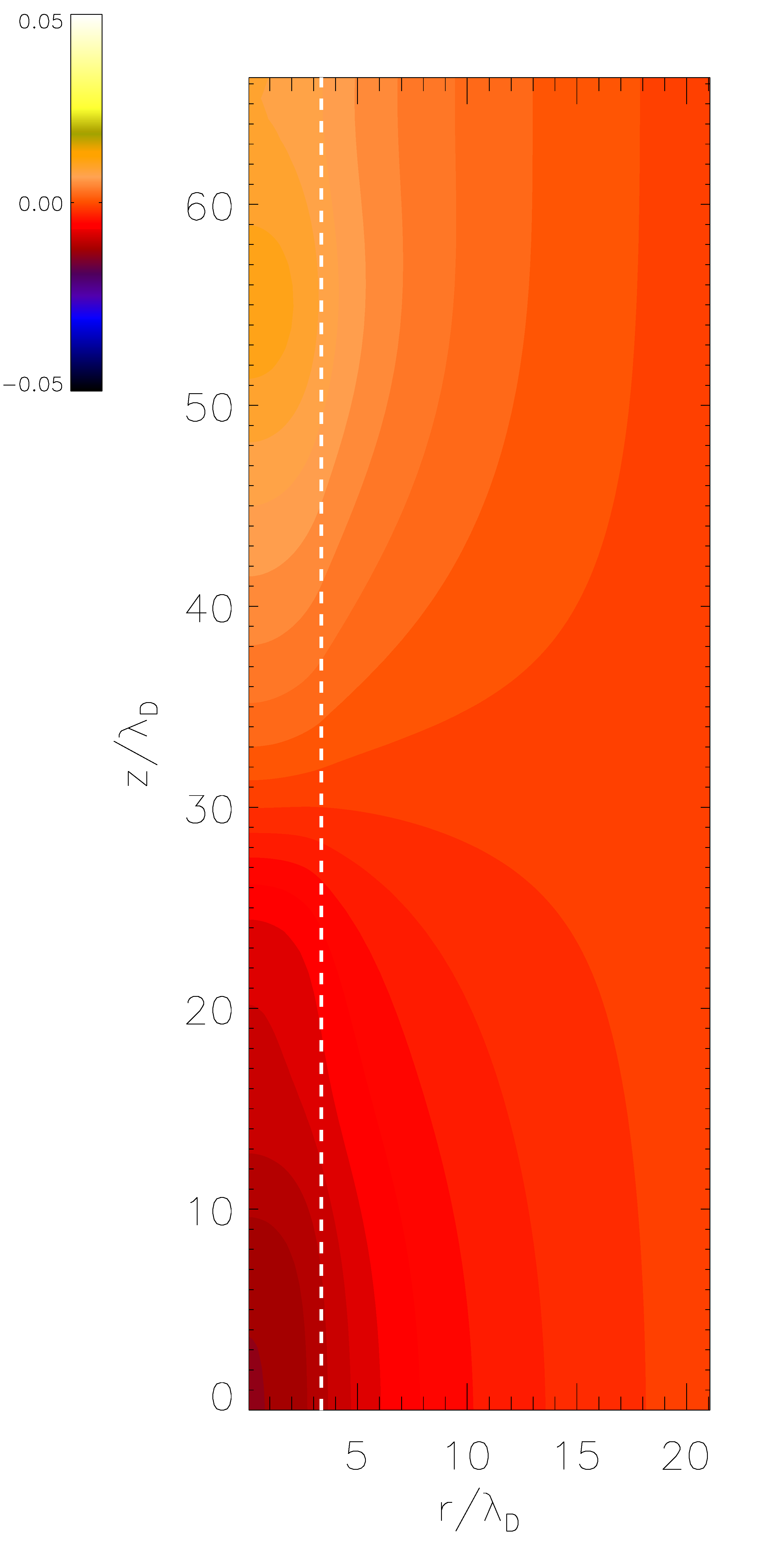}
  \end{minipage}
\caption{Contour plots of $\phi(r,z,t)$, scaled to $k_B T/e$, in the plane $(r,z)$ at different instants of time during the simulation of EAWs excitation. \textbf{The top row corresponds to the ramping up phase of the driver amplitude (at the time instants $t=17025, 17050, 17075, 17100$, from left to right), while the bottom row is focused on the ramping down phase (at the time instants $t=22900, 22925, 22950, 22975$, from left to right).} The vertical white-dashed line in each plot represents the radius $R_p$ of the plasma column.}
\label{fig:snap_movie}
\end{figure*}
In order to excite TG and EA fluctuations in physical conditions close to the ones discussed by \citet{anderegg2009electron,anderegg2009waveparticle}, we consider a plasma of single-ionized magnesium with density $n \simeq 1.5\times 10^7$cm$^{-3}$ and temperature $k_B T\simeq 0.5$eV. In these conditions we have plasma frequency $f_p=\omega_p/2\pi = 165$kHz, Debye length $\lambda_D\simeq 0.136$cm, thermal speed $v_{th}=\lambda_D\omega_p = 140$cm/ms and plasma parameter $g=1/n\lambda_D^3=2.7\times10^{-5}$. The wall radius and the plasma length and radius are set as $R_w=2.86$cm, $L_p=9$cm, $R_p=0.45$cm. We remind the reader that a sketch of the experimental ion trap is shown in Fig. \ref{fig:picexp} and the numerical box in  physical space for these simulations is depicted in Fig. \ref{fig:picnum}.

\begin{table}
\begin{center}
\begin{tabular}{|c|cccccc|}
\colrule
%RUN &  $t_{max}[\omega_p^{-1}]$ &\; $t_1[\omega_p^{-1}]$ &\; $t_2[\omega_p^{-1}]$ &\; $\Delta t[\omega_p^{-1}]$ &\; $V_D[mv_{th}^2/e]$ &\; $\omega_D[\omega_p]$  \\
RUN &  $t_{max}$ &\; $t_1$ &\; $t_2$ &\; $\Delta t$ &\; $V_D$ &\; $\omega_D$  \\
\colrule
TGW &  $27000$ &\; $3720$ &\; $4680$ &\; $57$ &\; $0.017$ &\; $0.177$ \\
EAW &  $84100$ &\; $16000$ &\; $24000$ &\; $1000$ &\; $0.05$ &\; $0.0687$  \\
\colrule
\end{tabular}
\end{center}
\caption{Parameters of the numerical simulations of section \ref{labsim}.}
\label{tab:param}
\end{table}

The relevant driver parameters for the TGWs and EAWs simulations are listed in Table \ref{tab:param} and have been set to get successful excitation of TGWs (EAWs) 
by driving the plasma for $15$ ($100$) wave cycles, with frequency $f_D=29.1$kHz ($f_D=11.3$kHz) and amplitude $8.5$($25$mV).

\begin{figure*}
\includegraphics[width=0.8\textwidth]{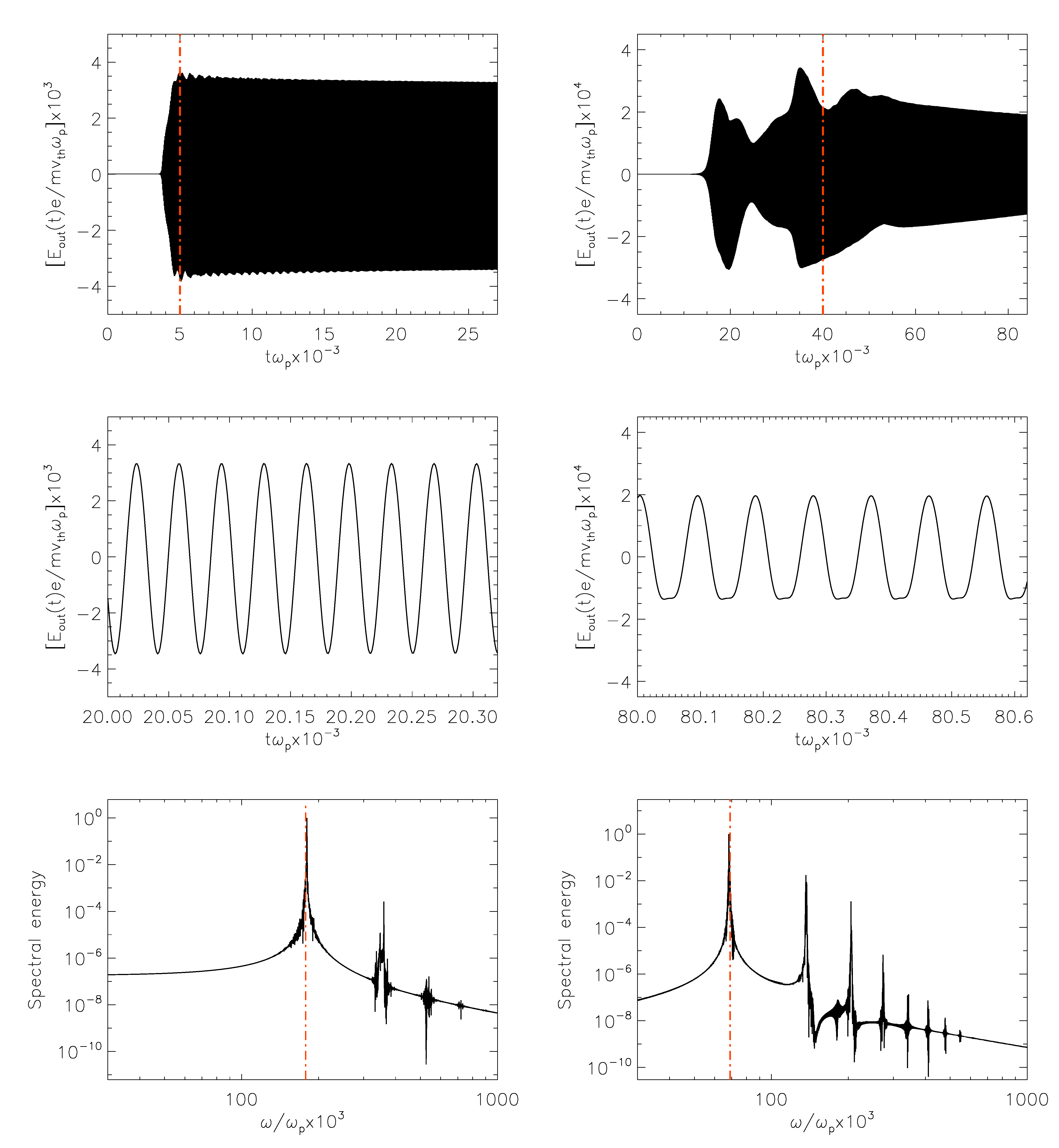}       % FIGURE 9
\caption{Top row: time evolution of the received signal $E_{out}$ for TG (left) and EA (right) fluctuations; 
here, the vertical dot-dashed lines indicate the time at which the external driver is turned off. Middle row: a zoom on few wave cycles of 
the signal $E_{out}$ for TG (left) and EA (right) fluctuations, taken in a time interval where the external driver is off. Bottom row: Spectral energy of TG 
(left) and EA (right) fluctuations, as a function of $\omega/\omega_p$; here, the vertical red-dashed line represents the driver frequency $\omega_D$. } \label{fig:TGEAmodes}
\end{figure*}

In Fig. \ref{fig:snap_movie}, we show how the fluctuating driving potential $\phi_{exc}$ applied to an electrode of length $\Delta z/2$ located at $r=R_w$ (see Fig. \ref{fig:picnum}) can trigger the fluctuations in the plasma column. Each panel in this figure reports the contour plot of $\phi(r,z,t)$ in the plane $(r,z)$ at different instants of time during the simulation, for the case of EAWs excitation. The top-row panels report four instants of time corresponding to the ramping-up phase of the driver amplitude: $t=17025, 17050, 17075, 17100$. Bottom-row panels are focused on the ramping-down phase at times: $t=22900, 22925, 22950, 22975$. The vertical white-dashed line in each plot represents the radius $R_p$ of the plasma column. The excitation of the EAWs by the external driver can be better appreciated in the movies we uploaded as the supplementary material, concerning the ramping up and down phases of the driver amplitude. As it can be seen, the external field makes the potential in the plasma column oscillate resonantly and the oscillations survive even after the driver has been turned off.

To make contact with experimental measurements, we focus on the signal $E_{out}$, defined as the radial electric field $E_r=-\partial\phi/\partial r$ calculated at the wall $r=R_w$, and integrated over the longitudinal length $\Delta z/2$ (see Fig. \ref{fig:picnum}):
\begin{equation}
\label{phi_out}
 E_{out}(t) = \frac{1}{\Delta z/2} \int_{0}^{\Delta z/2} dz \;E_r(R_w,z,t).
\end{equation}
The signal $E_{out}$ is an estimation of the signal measured in real experiments. Indeed, the radial electric field induces a surface charge on the wall electrode. This instantaneous charge on the electrode is proportional to the integral of the radial electric field over the area of the electrode, or —for an azimuthally symmetric mode— to the integral over the axial length of the electrode, that is, Eq. (\ref{phi_out}). Since the mode fluctuation is a temporally varying field, the surface charge on the electrode varies in time; hence, the current to supply this charge runs from ground to the electrode through, in general, a detecting and amplifying apparatus characterized by a certain impedance, that converts such a current in the observed voltage. Such a voltage is hence the detected wave signal, which is proportional to the wave amplitude at the location of the electrode. Finally, note also that the diagnostic based on Eq. (\ref{phi_out}) is not in conflict with the boundary conditions specified by Eqs. (\ref{eq:radial_BC}) and (\ref{eq:bc_phiv_rad}) since a vanishing potential does not imply a vanishing radial electric field.

The top row of Fig. \ref{fig:TGEAmodes} reports the time evolution of $E_{out}(t)$ for TG (left) and EA (right) fluctuations. In both plots, we notice that after the driver is turned off (red dot-dashed vertical lines in the plots), nearly stable oscillations are observed, even though the received EA signal clearly shows a more erratic time behavior with respect to the TG signal. A very similar behavior has been reported for the excitation of TG and EA fluctuations in laboratory experiments (see Figs. 4 and 5 of \citet{anderegg2009electron}). The middle row of the same figure shows a zoom of $E_{out}(t)$ on few wave cycles for TGWs (left) and EAWs (right), in the time interval where the driver potential has been already turned off. Here, it can be easily appreciated that, while TG oscillations display a sinusoidal form, the EA waveform departs from a regular sinusoid, suggesting the external driver has triggered the excitation of several wavenumbers. To highlight this point, in the bottom row of Fig. \ref{fig:TGEAmodes}, we report the spectral energy (i.e., the squared absolute value of the Fourier transform of $E_{out}(t)$), evaluated in the time interval where the driver is turned off, as a function of $\omega$ for both TGWs (left) and EAWs (right). In each panel, the black curve has been scaled to its maximum and the vertical red-dashed line corresponds to the driver frequency $\omega_D$. In both plots, the black curve is peaked around the driver frequency. For the TG fluctuations, the energy content of the second (third) harmonics is four (six) orders of magnitude smaller than that of the fundamental, thus implying the driving of quasi-monochromatic fluctuations. The small-amplitude high-order harmonics are likely due to the larger temperature and amplitude used in these simulations at variance with the ones reported in Sec. \ref{drivtest} where no high-order harmonics were recovered. For the case of EA fluctuations clearly the contribution of harmonics is non-negligible, resulting in the waveform displayed in the middle-right panel of Fig. \ref{fig:TGEAmodes}. The erratic behavior observed for EAWs and related to their non-monochromatic features reflects also the fact that the creation of the velocity plateau in the case of the EAW excitation is a rather complicated process, occurring in the core of the velocity distribution, where much more particles are in resonance with the wave, than in the case of the TGW excitation, which instead occurs in the tail of the velocity distribution. By looking at the spectral energy in different time windows (not explicitly shown here), we notice that higher harmonics are generated while the driver is still on, despite the driver frequency corresponds to that of the fundamental harmonic (red dot-dashed vertical line in Fig. \ref{fig:TGEAmodes}). This can be understood since the process is nonlinear and the driver is maintained on for about 90 wave periods, computed with the fundamental harmonic. There is hence enough time to generate secondary harmonics while the driver is still on. These harmonics are rather stable and persist without significant changes in the analyzed time windows. Future analysis, beyond the scope of the present work, will focus on possible interactions between these harmonics.

\begin{figure}
\includegraphics[width=220pt]{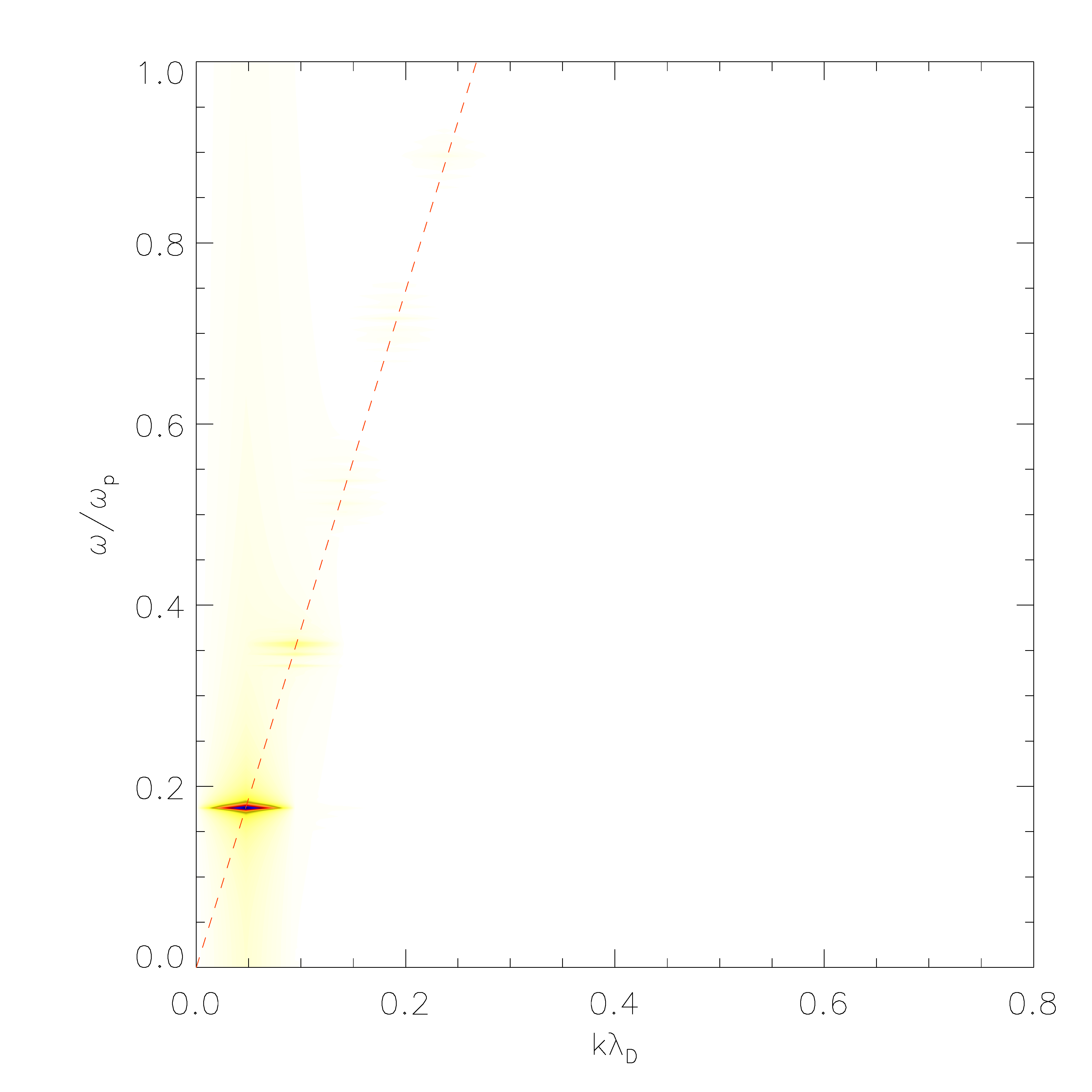}   % FIGURE 10
\includegraphics[width=220pt]{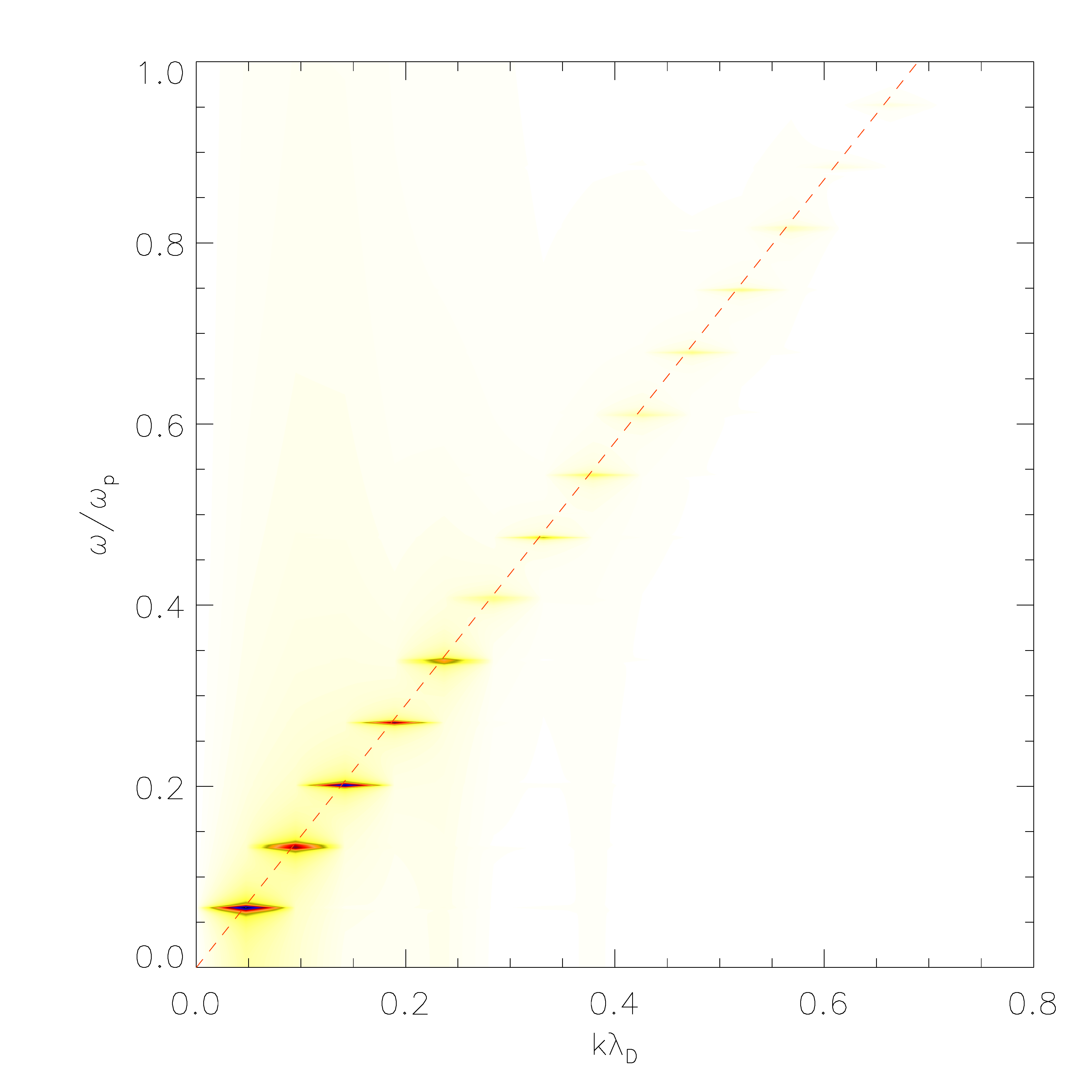}
\caption{$\omega$-$k$ Fourier spectrum of the numerical signals from TG (top) and EA (bottom) simulations; in both plots the 
red-dot-dashed lines represents the curve $\omega=kv_\phi=k\omega_D/k_1$.} \label{fig:wk}
\end{figure}

\begin{figure*}
\includegraphics[width=\textwidth]{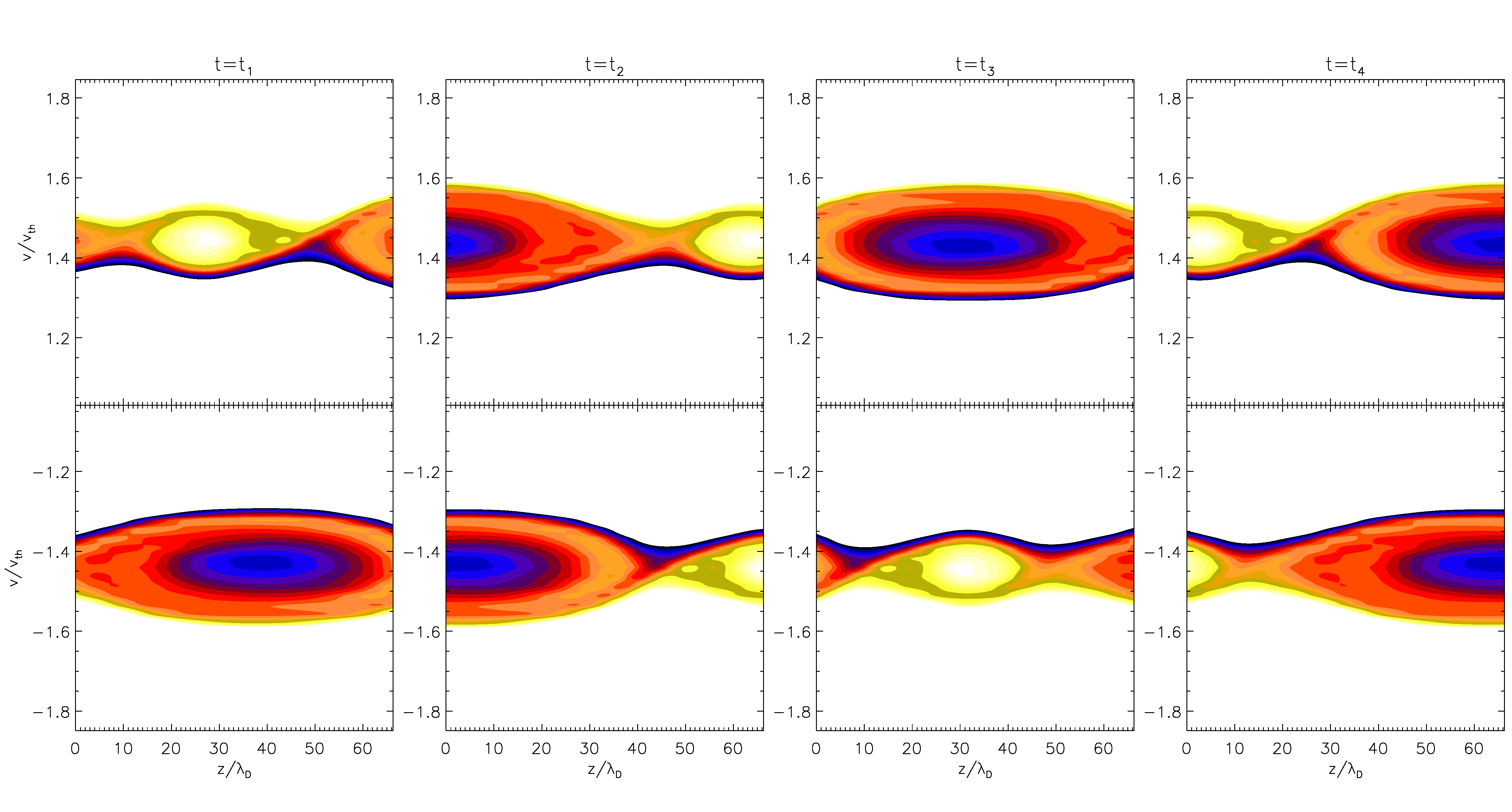}  % FIGURE 11
\caption{$z$-$v$ phase space portrait of the trapped particle distribution function at $r=0$ at four instants of time (after the external driver has been turned off), for the simulation of the EAWs excitation. Vortical structures, typical signature of the presence of a trapped particle population, move in circle in phase space, being reflected (i. e. changing sign of their mean speed) at each end of the plasma column, at $z=0$ and $z=L_p$.} \label{fig:VDF}
\end{figure*}

As the numerical simulations allow to study both the spatial and temporal variations of the electric potential, in Fig. \ref{fig:wk} we report the Fourier $\omega-k$ spectrum of the numerical signal $\phi(r=0,z,t)$ inside the plasma column (i. e. at $r=0$), for the case of TG (top) and EA (bottom) excitation, evaluated in a time interval when the external driver has been turned off. In these two plots, for visualization purposes, the spectra have been scaled in such a way the amplitude of the more energetic peak (for both cases the one corresponding to the fundamental wavenumber) is equal to unity (in dimensionless units). In both plots, the red-dot-dashed lines represent $kv_\phi$, where $v_\phi=\omega_D/k_1$, $k_1=\pi/L_p$ being the fundamental longitudinal wavenumber. As it is clear from these spectra, for the case of TG waves the driving fluctuating potential applied to the external confining cylinder has triggered quite monochromatic fluctuations; on the other hand, for the case of the EA waves, the resulting oscillations are composed of several wavenumbers, excited along a straight line at the driver frequency (with the same phase velocity). In both cases, however, the driver applied to the external cylinder (at $r=R_w$) is able to excite fluctuations and to trigger the propagation of waves into the plasma column.

Finally, in Fig. \ref{fig:VDF} we display the $z$-$v$ phase space distribution function (evaluated at $r=0$ and for $0\leq z\leq L_p$) of particles trapped in the wave potential well, for the case of the EA excitation. The four columns in this figure (from left to right) correspond to four different time instants \textbf{($t_1=84025, t_2=84050, t_3=84075, t_4=84100$)} in the simulations, almost covering a wave cycle. In the top (bottom) row, we show the trapping region at positive (negative) velocities, around the phase speed of the driver. This feature is peculiar of these nonlinear waves, which can exist only in the presence of trapped particles. We notice a main vortical structure and a secondary smaller island, typical signature of particle trapping. These vortices move in circle in phase space, this corresponding to a periodic motion along the 
$z$-direction, with specular reflection at each end of the plasma column. Indeed, at $t=t_1$, the main vortex is almost entirely in the negative velocities portion of phase space, traveling from the right to the left. Then, at $t=t_2$, it appears in the positive part of the velocity domain, traveling from the left to the right and, at $t=t_3$ it is located in the middle of the positive velocity domain. Once the vortex arrived at the right end ($t=t_4$), it is reflected (changing sign of its mean speed) and goes back to the negative velocity range.

%%%%%%%%%%%%%%%%%%%%%%%%%%%%%%%%%%%%%
\section{Summary and conclusions}\label{sect:concl}
%%%%%%%%%%%%%%%%%%%%%%%%%%%%%%%%%%%%%
In this paper, a newly developed Eulerian code has been presented, which integrates the (azimuthally homogeneous) drift-kinetic Poisson equations in 2D-1V phase space (two dimensions in physical space and one dimension in velocity space) in cylindrical coordinates, suited for the description of the kinetic dynamics of a non-neutral plasma. The algorithm has been specifically designed to reproduce the excitation of nearly acoustic plasma waves (Trivelpiece-Gould and Electron Acoustic waves) in a Penning-Malmberg apparatus, in typical conditions of a real laboratory experiment.  

Periodic boundary conditions are implemented in the longitudinal direction $z$, and the total length of the numerical $z$-box is twice the actual length of the plasma column; moreover, the wave electric field is imposed to be null at each end of the plasma column. In these conditions, we are able to model particle and wave reflection at each axial end of the plasma column. The radial boundary conditions are set in such a way we can model the wave launching process in the fashion of a real experiment that is through an external fluctuating potential applied to an electrode of finite axial length, located on the external conducting cylinder, at a distance $R_w$ from the main axis of the confining trap. The equilibrium plasma density has a top-hat profile with a sharp discontinuity at the plasma radius $R_p$ where the density changes from $n_0$ to $0$. This approximation does not model the plasma sheath, whose size is only a few Debye lengths, corresponding to the exponential decrease in density. This choice is fully justified in conditions where the Debye length is small compared to the radius of the plasma column. Previous theoretical works on modes in a long column with a finite Debye length drop in density also indicate that results are unchanged for modes with sufficiently low-order modes, while some differences appear in higher-order modes for which the radial wavelength becomes comparable to the Debye scale. Our simulations of cold plasma (Secs. \ref{ivptest} and \ref{drivtest}) well satisfy the above condition, since $R_p\sim28\lambda_D$. On the other hand, simulations of warm plasma (Sec. \ref{labsim}) do not have such a large separation of scales since $R_p\sim 3 \lambda_D$. However, in both cases, we focus on low-order modes, thus likely justifying the adoption of a sharp drop in the equilibrium plasma density. We will extend our results to the case of a finite plasma density falloff in future work.

The code has been first tested by (i) launching linear Trivelpiece-Gould waves in the plasma column as an initial value problem, and (ii) exciting linear TG fluctuations through an external driving potential applied to the radial wall of the conducting cylinder. A detailed comparison with theoretical expectations suggests that the algorithm is able to model correctly the excitation of both linear TG modes. Therefore, the excitation of TG and EA fluctuations has been triggered through an external potential reproducing closely the phenomenology of laboratory experiments performed at University of California at San Diego~\citep{anderegg2009electron,anderegg2009waveparticle}. We remark that the Eulerian approach adopted here guarantees a rather clean description of dynamics in linear and nonlinear cases. A PIC approach would have suffered from the intrinsic thermal noise that could have caused a deprecation of the signal unless massively increasing the number of particles per cell \citep{camporeale2011dissipation,bacchini2022kinetic}. Indeed, to make contact with linear TGWs predictions, the amplitude of the fluctuations in the numerical simulations must be very low and the noise level introduced in PIC simulations risks to completely mask the signal. On the other hand, EAWs are associated with a peculiar structure in velocity space that, again, requires a noiseless description of the full phase space (see, e.g., Refs.\citep{pezzi2016collrec,pezzi2017colliding,pezzi2018velocityspace,cerri2018dual}).

Our numerical code is parallelized with the shared-memory directives OpenMP. This parallelization is rather efficient and allows us to perform a typical run for the excitation of TGWs (see Table \ref{tab:param}) in about 18 h (speed-up of 15 with respect to the serial execution) on the NEWTON HPC cluster at the University of Calabria, equipped with 128GB of RAM memory and 20 computing cores per node, achieving a speed-up of about 15. Estimating the efficiency of the parallelization, which we do not test against different machines or by changing the number of cores, is beyond the scope of this paper and will be addressed in the future. In future work, we also plan to include in the algorithm the effect of Coulomb collisions, by implementing the Dougherty nonlinear collisional operator\citep{dougherty1964model,dougherty1967model2}. We also intend to include the azimuthal dependence, thus moving the numerical computation to a full 6D phase-space geometry. Such a configuration would likely require the adoption of a more advanced parallelization strategy based on a hybrid MPI/OpenMP approach.

This Eulerian drift-kinetic Poisson code can be extremely useful as a support for the interpretation of the experimental results obtained in Penning-Malmberg machines, confining non-neutral plasmas, as it allows for a relevant point-to-point comparison between numerical results and laboratory measurements.

\section*{Supplementary material}
See the supplementary material for the detailed illustration of the launching process due to the external forcing. "movie-up-LR.avi" and "movie-down-LR.avi" ,respectively, correspond to the ramping up and down phases of the driver amplitude.

% Figures should be put into the text as floats. 
% Use the graphics or graphicx packages (distributed with LaTeX2e).
% Put the label that you will use with \ref{} command in the braces of the \label{} command.
%
% \begin{figure}
% \end{figure}

% \begin{table}
% \end{tabular}
% \end{table}

\begin{acknowledgments}
The kinetic simulations discussed in the present paper have been run on the NEWTON cluster at the University of Calabria by means of the PEnning Trap wavE simulatoR (PETER) numerical code developed by F.V. and O.P. The research of T.M. O'Neil was supported by NSF Grant No. PHY2106332 and U.S. Department of
Energy Grant No. DE-SC0018236.
\end{acknowledgments}

\section*{Data Availability Statement}
The data that support the findings of this study are available from the corresponding author upon reasonable request.

\appendix*

\section{ Reflection mixing}

For the geometric parameters assumed in the simulations and the case of a cold plasma, Fig. \ref{fig:omek} shows that the mode $(3, 2)$ is nearly degenerate with the mode $(1, 1$). In this case, reflection mixing adds a small component of the $(3, 2)$ mode to the $(1, 1)$ mode, and Eq. (\ref{eq:amplitude_ratio}) provides an estimate of the relative size of the admixture. The purpose of this appendix is to explain the estimate in more detail following the analysis of \citet{anderson2011degenerate}. 

The simple model developed in this paper starts by finding linear solutions to the coupled Vlasov-Poisson equations for an infinitely long plasma column. For a generic solution characterized by a single axial wave number $k$, Eq. (\ref{eq:phi_continuity}) determines a series of values $k_{\perp n,m}(k)$, while the dispersion relation in Eq. (\ref{eq:dispersion relation}) provides the complex frequency $\omega=\omega(k,m)$. For the simple model adopted here, the eigenmodes for the finite length plasma column are then determined by implementing the periodic boundary condition $k_{n}=(2\pi n)/(2L_{p})=\pi n/L_{p}$.

However, this simple boundary condition omits reflection mixing. Because of this mixing, a true eigenmode for the finite-length column consists of a superposition of the simple model eigenmodes with various values of $k$. For the true eigenmodes, $k$ is not a good quantum number, because the plasma equilibrium is not invariant under translations in $z$. The frequency $\omega$ is still a good quantum number because the equilibrium is invariant under translations in time. Thus, for the true eigenfunctions, it is useful to turn the dispersion relation $\omega=\omega(k,m)$ around and to think of $k=k(\omega,m)$.

Since the plasma equilibrium is invariant under reflection about its midplane, the true eigenmodes must have either even parity under such reflection or odd parity; that is, the true eigenmodes contain either mixtures of simple model eigenmodes for even $n$, or separately, for odd $n$. We will consider the case of odd parity, having in mind the near degeneracy of the simple model modes  $(1, 1)$ and $(3, 2)$.  

Assuming that only two odd modes participate significantly in the mixing, the boundary condition at the plasma end  takes the form
\begin{equation}\label{eq:app_1}
    A_{m}(\omega)k(\omega,m)\cot{\left[\dfrac{k(\omega,m)L_{p}}{2}\right]}= \textit{O}\left(\dfrac{R_{w}}{L_{p}^{2}}\right)A_{m'}(\omega)
\end{equation}
\begin{equation}\label{eq:app_2}
    A_{m'}(\omega)k(\omega,m')\cot{\left[\dfrac{k(\omega,m')L_{p}}{2}\right]}= \textit{O}\left(\dfrac{R_{w}}{L_{p}^{2}}\right)A_{m}(\omega)
\end{equation}
where the right-hand sides have been approximated by order of magnitude estimates \citep{anderson2011degenerate}. The function $\cot{[k(\omega,m)L_{p}/2]}$ vanishes when $\omega$ is chosen so that $k(\omega,m)L_{p}=\pi n$, where $n$ is an odd integer, and the other cotangent vanishes when $k(\omega,m')L_{p}=\pi n'$. We have in mind here that $(n, m)=(1, 1)$ and $(n', m')= (3, 2)$.  

First, we consider the case where $k(\omega,m)L_{p}$ is very nearly equal to $\pi n$ and the inequality
\begin{equation}
k(\omega,m)\cot{[k(\omega,m)L_{p}/2] \ll \textit{O}}(R_{w}/L_{p}^{2})
\end{equation} 
is satisfied. Equation (\ref{eq:app_1}) then  implies the inequality $A_{m}(\omega)\gg A_{m'}(\omega)$, so $A_{m}(\omega_{n,m})\equiv A_{n,m}$ is dominant. By hypothesis modes $(n, m)$ and $(n', m')$ are nearly degenerate, so the cotangent function in Eq. (\ref{eq:app_2}) also is approximately null; hence,
\begin{equation}\label{eq:app_3}
    k(\omega,m')\cot{\left[\dfrac{k(\omega,m')L_{p}}{2}\right]}\approx \dfrac{\pi n'}{L_{p}}\dfrac{L_{p}\delta k_{n'}}{2}
\end{equation}
where
\begin{equation}\label{eq:app_4}
    \delta k_{n'}=\dfrac{(\omega_{n,m}-\omega_{n',m')}}{\dfrac{\partial \omega_{n',m'}}{\partial k_{n'}}}\simeq \dfrac{(\omega_{n,m}-\omega_{n',m'})k_{\perp n'm'}}{\omega_{p}}
\end{equation}
By combining Eqs. (\ref{eq:app_2})-(\ref{eq:app_4}), we finally get
\begin{eqnarray}
    \dfrac{A_{n',m'}}{A_{n,m}} &&= \dfrac{R_{w}}{L_{p}^{2}n'\pi k_{\perp n',m'}}\dfrac{\omega_{p}}{|{\omega_{n,m}-\omega_{n',m'}|}} \nonumber \\
    && \sim \dfrac{R_{w}R_{p}}{L_{p}^{2}n' \pi(1+3m')}\dfrac{\omega_{p}}{|{\omega_{n,m}-\omega_{n',m'}}|}
\end{eqnarray}
where in the last step we used the approximation $k_{\perp n',m'}\sim (1+3m')$. This final form is equivalent to Eq. (\ref{eq:amplitude_ratio}). Note that, for the case of a cold plasma, the mode frequency $\omega_{n,m}$ is simply equal to $\omega_{p}(n,m)$, and the Landau damping rate is zero. Moreover, Fig. \ref{fig:omek} suggests that $\omega_{1,1}=\omega_{p}(1,1)$ and $\omega_{3,2}=\omega_{p}(3,2)$ are relatively close in value; indeed, $(\omega_{3,2}-\omega_{1,1})/\omega_{p} \sim 10^{-2}$. For such a case, the admixture level is rather small, being $A_{n',m'}/A_{n,m}\sim 10^{-2}$. 

\bibliography{BIBLIO-living}

\end{document}